\documentclass[%
reprint,
superscriptaddress,
amsmath,amssymb,
aps,
rmp]{revtex4-1}
\usepackage{graphicx}
\usepackage{dcolumn}
\usepackage{bm}
\usepackage{physics}
\usepackage{braket}
\usepackage{amssymb}
\usepackage{placeins}
\usepackage{array}

\usepackage{graphicx}

\usepackage{amssymb}
\usepackage{amsthm}

\usepackage{listings}
\newcommand{\inline}[1]{%
  \lstinline[columns=fixed]{#1}}

\RequirePackage[colorlinks=true,linkcolor=blue,urlcolor=blue,hyperfootnotes=false,citecolor=black]{hyperref}
\usepackage{braket}
\usepackage{physics}

\begin{document}

\title{QuCAT: Quantum Circuit Analyzer Tool in Python}

\author{Mario F. Gely}
\affiliation{%
Kavli Institute of NanoScience, Delft University of Technology, PO Box 5046, 2600 GA, Delft, The Netherlands.
}

\author{Gary A. Steele}
\affiliation{%
Kavli Institute of NanoScience, Delft University of Technology, PO Box 5046, 2600 GA, Delft, The Netherlands.
}

\begin{abstract}
Quantum circuits constructed from Josephson junctions and superconducting electronics are key to many quantum computing and quantum optics applications.
Designing these circuits involves calculating the Hamiltonian describing their quantum behavior.
Here we present QuCAT, or ``Quantum Circuit Analyzer Tool", an open-source framework to help in this task.
This open-source Python library features an intuitive graphical or programmatical interface to create circuits, the ability to compute their Hamiltonian, and a set of complimentary functionalities such as calculating dissipation rates or visualizing current flow in the circuit.
\end{abstract}

\date{\today}

\maketitle
\tableofcontents

\section{Introduction}
Quantum circuits, constructed from superconducting electronics and involving one or more Josephson junctions, have steadily gained prominence in experimental and theoretical physics over the past twenty years.
Foremost, they are one of the most successful platforms in the quest to build a quantum computer~\cite{Devoret2013}.
The control that can be gained over their quantum state, and the flexibility in their design have also made these circuits an excellent test-bed to probe fundamental quantum effects~\cite{gu2017microwave}.
They can also be coupled to other systems, such as atoms, spins, acoustic vibrations or mechanical oscillators, acting as a tool to measure and manipulate these systems at a quantum level~\cite{xiang2013hybrid}.
%

Any application mentioned above generally translates to a desired Hamiltonian, which governs the physics of the circuit.
The task of the quantum circuit designer is to determine which circuit components to use, how to inter-connect them, and calculate the corresponding Hamiltonian~\cite{vool2017introduction,nigg_black-box_2012}.
%
%
Performing this task analytically can be time consuming or even challenging.
%

Here we present QuCAT, which stands for ``Quantum Circuit Analyzer Tool", an open-source Python framework to help in analyzing and understanding quantum circuits.
We provide an easy interface to create and visualize circuits, either programmatically or through a graphical user interface.
A Hamiltonian can then be generated for further analysis in QuTiP~\cite{johansson2012qutip,johansson2013qutip}.
The current version of QuCAT supports quantization in the basis of normal modes of the linear circuit~\cite{nigg_black-box_2012}, making it suited for the analysis of weakly anharmonic circuits with small losses.
The properties of these modes: their frequency, dissipation rates, anharmonicity and cross-Kerr couplings can be directly calculated.
The user can also visualize the current flows in the circuit associated with each normal mode.
The library covers lumped element circuits featuring an arbitrary number of Josephson junctions, inductors, capacitors and resistors.
Through equivalent lumped element circuits, certain distributed elements such as waveguide resonators can also be analyzed (see Sec.~\ref{sec:mmusc}).
The software relies on the symbolic manipulation of the circuits equations, making it reliable even for vastly different circuits and parameters.
It also results in efficient parameter sweeps, as analytical manipulations need not be repeated for different circuit parameters.
In a few seconds, circuits featuring 10 nodes (or degrees of freedom), corresponding to between 10 and 30 circuit elements can be simulated.

In the main section of this article, we cover the functionalities of the software.
We start by showing how to create circuits, first using the graphical user interface, then programmatically.
We then demonstrate how to generate the corresponding Hamiltonian.
Lastly, we show how to extract the characteristics of the circuit modes: frequencies, dissipation, anharmonicity and cross-Kerr coupling and present a tool to visualize these modes. 
This main section will feature as an example the standard circuit of a transmon qubit coupled to a resonator~\cite{Koch2007}.
In the appendices, we will first use QuCAT to analyze some recent experiments: a tuneable coupler~\cite{kounalakis2018tuneable}, a multi-mode ultra-strong coupling circuit~\cite{bosman_multi-mode_2017}, a microwave optomechanics circuit~\cite{ockeloen2016low} and a Josephson-ring based qubit~\cite{roy2017implementation}.
We then provide an overview of the circuit quantization method used and the algorithmic methods which implement it.
The limitations of these methods regarding weak anharmonicity and circuit size will then be presented.
Finally we will explain how to install QuCAT and we provide a summary of all its functions.
More tutorials and examples are available on the QuCAT website \url{https://qucat.org/}.

\section{Circuit construction}
\begin{figure}[ht!]
\includegraphics[width=0.5\textwidth]{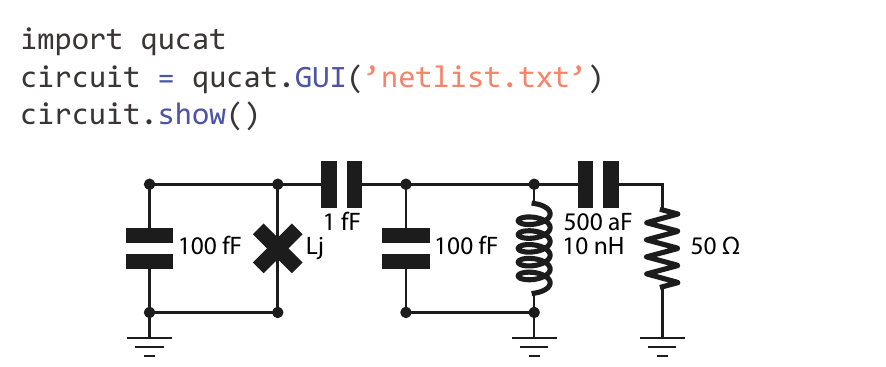}
\caption{
\textbf{Construction of a circuit: }
code and output.
The circuit used as an example in this section comprises of a transmon qubit on the left, coupled through a 1 fF capacitor to an LC-oscillator.
Dissipation arises from the capacitive coupling of the LC-oscillator to a 50 $\Omega$ resistor on the right.
After importing the \inline{qucat} package, the \inline{circuit} object is created manually through a graphical user interface (GUI) opened after calling \inline{qucat.GUI("netlist.txt")}.
All information necessary to construct the circuit is stored in the text file \inline{netlist.txt}.
After closing the GUI, this information is also stored in the variable \inline{circuit}.
The \inline{show} method finally displays the circuit.
}
\label{fig:1}
\end{figure}
Any use of QuCAT will start with importing the \inline{qucat} library
\begin{lstlisting}
import qucat
\end{lstlisting}
One should then create a circuit.
These are named \inline{Qcircuit}, short for ``quantum circuit" in QuCAT.
There are two ways of creating a \inline{Qcircuit}: using the graphical user interface (GUI), or programmatically.

\subsection{Creating a circuit with the GUI}

We first cover how to create a circuit with the GUI.
This is done through this command 
\begin{lstlisting}
circuit = qucat.GUI('netlist.txt')
\end{lstlisting}
which opens the GUI.
The GUI will appear as a separate window, which will block the execution of the rest of the Python script until the window is closed.
The user can drag-in and drop capacitors, inductors, resistors or Josephson junctions, or grounds.
These components can then be inter-connected with wires.
Each change made to the circuit will be automatically be saved in the \inline{'netlist.txt'} file.
After closing the GUI, the \inline{Qcircuit} object will be stored in the variable named \inline{circuit} which we will use for further analysis.

\subsection{Creating a circuit programmatically}
Alternatively, one can create a circuit with only Python code.
This is done by creating a list of circuit components with the functions \inline{J}, \inline{L}, \inline{C} and \inline{R} for junctions, inductors, capacitors and resistors respectively.
For the circuit of Fig.~\ref{fig:1}:
\begin{lstlisting}
circuit_components = [
qucat.C(0,1,100e-15), # transmon
qucat.J(0,1,'Lj'),
qucat.C(0,2,100e-15), # resonator
qucat.L(0,2,10e-9),
qucat.C(1,2,1e-15), # coupling capacitor
qucat.C(2,3,0.5e-15), # ext. coupl. cap.
qucat.R(3,0,50) # 50 Ohm load
]
\end{lstlisting}
All circuit components take as first two argument integers referring to the negative and positive node of the circuit components.
Here 0 corresponds to the ground node for example.
%
%
The third argument is either a float giving the component a value, or a string which labels the component parameter to be specified later.
Doing the latter avoids performing the computationally expensive initialization of the \inline{Qcircuit} object multiple times when sweeping a parameter.
By default, junctions are parametrized by their Josephson inductance $L_j = \phi_0^2/E_j$
where $\phi_0 = \hbar/2e$ is the reduced flux quantum, 
and $E_j$ (in Joules) is the Josephson energy.

Once the list of components is built, we can create a \inline{Qcircuit} object via the \inline{Network} function
\begin{lstlisting}
circuit = Network(circuit_components)
\end{lstlisting}
as with a construction via the GUI, the \inline{Qcircuit} object will be stored in the variable named \inline{circuit} which we will use for further analysis.

\section{Generating a Hamiltonian}
\begin{figure}[ht!]
\includegraphics[width=0.5\textwidth]{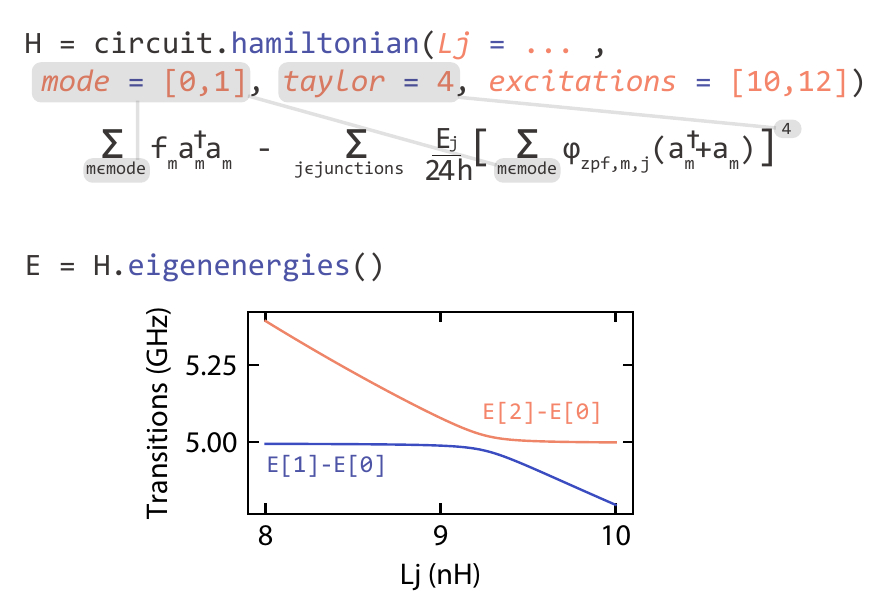}
\caption{
\textbf{Hamiltonian generation}
is done by applying the \inline{hamiltonian} method to the \inline{circuit} variable defined in Fig.~\ref{fig:1}.
The Hamiltonian is expressed in the basis of circuit normal modes $m$ with frequencies $f_m = \omega_m/2\pi$, annihilation operators $\hat{a}_m$, and zero-point phase fluctuations $\varphi_{\text{zpf},m,j}$ across junction $j$ with Josephson energy $E_j$.
The junction non-linearities are expressed through a Taylor expansion of the cosine potentials, where the user chooses the degree of Taylor expansion.
The other arguments are the list of modes to include, the number of excitations to consider for each of these modes, and any unspecified component value, here \inline{Lj}.
The returned Hamiltonian is a QuTiP object, giving the user access to an extensive set of tools for further analysis~\cite{johansson2012qutip,johansson2013qutip}.
As an example, we compute the eigenenergies of the Hamiltonian, and plot the two first transition frequencies, as a function of \inline{Lj}.
}
\label{fig:4}
\end{figure}
The Hamiltonian of a Josephson circuit is given by 
\begin{equation}
\hat{H} = \sum_{m} \hbar\omega_m\hat{a}_m^\dagger\hat{a}_m + \sum_j\sum_{n\ge 2}E_j\frac{(-1)^{n+1}}{(2n)!}\hat \varphi_j^{2n}\ .
\label{eq:hamiltonian_taylor}
\end{equation}
It is written in the basis of its normal modes.
These have an angular frequency $\omega_m$ and we write the operator which creates (annihilates) photons in the mode $\hat a_m^\dagger$ ($\hat a_m$).
The cosine potential of each Josephson junction $j$ with Josephson energy $E_j$ has been Taylor expanded to order $n$ for small values of its phase fluctuations $\hat \varphi_j$ across it.
The phase fluctuations are a function of the annihilation and creation operators of the modes $\hat\varphi_j=\sum_{m}\varphi_{\text{zpf},m,j}(\hat{a}_m^\dagger+\hat{a}_m)$.
For a detailed derivation of this Hamiltonian, and the method used to obtain its parameters, see Sec.~\ref{sec:circuit_quantization_overview}.

There are three different parameters that the user should fix
\begin{enumerate}
\item the set of modes to include
\item for each of these modes, the number of excitations to consider
\item the order of the Taylor expansion.
\end{enumerate}
The more modes and excitations are included, and the higher Taylor expansion order, the more faithful the Hamiltonian will be to physical reality.
The resulting increase in Hilbert space size will however make it more computationally expensive to perform further calculations.
Typically, larger degrees of anharmonicity require a larger Hilbert space, with a fundamental limitation on the maximum anharmonicity due to the choice of basis.
We expand on these topics in Sec.~\ref{sec:high_anharmonicity}.

Such a Hamiltonian is generated through the method \inline{hamiltonian}. 
More specifically, this function returns a QuTiP object~\cite{johansson2012qutip,johansson2013qutip}, enabling an easy treatment of the Hamiltonian.
All QuCAT functions use units of Hertz, so the function is actually returning $\hat H/h$.

As an example, we generate a Hamiltonian for the circuit of Fig.~\ref{fig:1} at different values of the Josephson inductance and use QuTiP to diagonalize it and obtain the eigen-frequencies of the system.
For a Josephson inductance of $8$ nH this is achieved through the commands
\begin{lstlisting}
H = circuit.hamiltonian(
        modes = [0,1],
        excitations = [10,12],
        taylor = 4,
        Lj = 8e-9)        
E = H.eigenenergies() # Eigenenergies (here in units of frequency) using the QuTiP function eigenenergies
\end{lstlisting}
With \inline{modes = [0,1]}, we are specifying that we wish to consider the first and second modes of the circuit.
Modes are numbered with increasing frequency, so here we are selecting the two lowest frequency modes of the circuit.
With \inline{excitations = [10,12]}, we specify that for mode \inline{0} (\inline{1}) we wish to consider \inline{10} (\inline{12}) excitations.
With \inline{taylor = 4}, we are specifying that we wish to expand the cosine potential to fourth order, this is the lowest order which will give an anharmonic behavior.
The unspecified Josephson inductance must now be fixed through a keyword argument \inline{Lj = 8e-9}. 
Doing so avoids initializing the \inline{Qcircuit} objects multiple times during parameter sweeps, as initialization is the most computationally expensive task.
We calculate these energies with different values of the Josephson inductance, and the first two transition frequencies are plotted in Fig.~\ref{fig:4}, showing the typical avoided crossing seen in a coupled qubit-resonator system.

\section{Mode frequencies, dissipation rates, anharmonicities and cross-Kerr couplings}
\begin{figure}[ht!]
\includegraphics[width=0.5\textwidth]{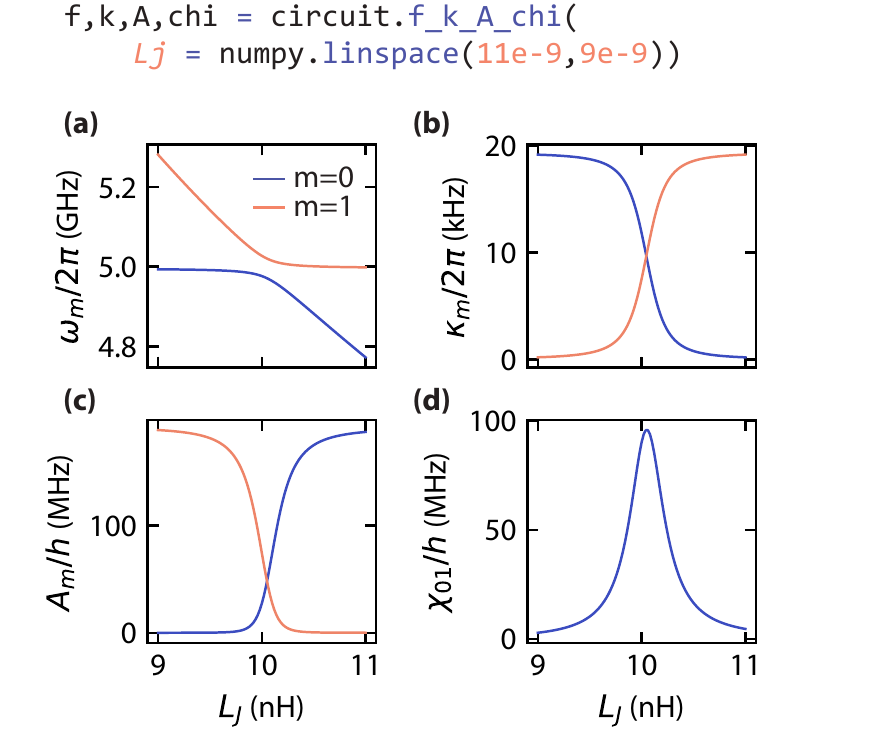}
\caption{
\textbf{Extracting eigenfrequencies, loss-rates, anharmonicities, and cross-Kerr couplings.}
We apply the \inline{f_k_A_chi} method to \inline{circuit} defined in Fig.~\ref{fig:1} to obtain a list of eigenfrequencies (\inline{f}), loss-rates (\inline{k}), anharmonicities (\inline{A}), and cross-Kerr couplings (\inline{chi}), for all the normal modes of the circuit.
There is one unspecified variable in the circuit, the Josephson inductance $L_j$, which is here specified with a list of values.
In \textbf{(a)}, we plot the eigenfrequencies of the two first modes \inline{f[0]} and \inline{f[1]}.
In \textbf{(b)}, we plot the loss-rates of the same modes \inline{k[0]} and \inline{k[1]}, and in \textbf{(c)} their anharmonicities \inline{A[0]} and \inline{A[1]}.
In \textbf{(d)}, we plot the cross-Kerr coupling between modes \inline{0} and \inline{1}: \inline{chi[0,1]}.
}
\label{fig:2}
\end{figure}
QuCAT can also return the parameters of the (already diagonal) Hamiltonian in first-order perturbation theory
\begin{equation}
\begin{split}
\hat{H} = \sum_m\sum_{n\ne m} (\hbar\omega_m-A_m-\frac{\chi_{mn}}{2})\hat{a}_m^\dagger\hat{a}_m \\
-\frac{A_m}{2}\hat{a}_m^\dagger\hat{a}_m^\dagger\hat{a}_m\hat{a}_m -\chi_{mn}\hat{a}_m^\dagger\hat{a}_m\hat{a}_n^\dagger\hat{a}_n
\label{eq:hamiltonian_first_order_maintext}
\end{split}
\end{equation}
valid for weak anharmonicity $\chi_{mn},A_m\ll \omega_m$.
The physics of this Hamiltonian can be understood by considering that an excitation of one of the circuit modes may lead to current traversing a Josephson junction. 
This will change the effective inductance of the junction, hence changing its own mode frequency, as well as the mode frequencies of all other modes.
This is quantified through the anharmonicity or self-Kerr $A_m$ and cross-Kerr $\chi_{mn}$ respectively.
When no mode is excited, vacuum-fluctuations in current through the junction give rise to shifted mode energies $\hbar\omega_m-A_m-\sum_n\chi_{mn}/2$.
In a circuit featuring resistors, these anharmonic modes will be dissipative.
A mode $m$ will lose energy at a rate $\kappa_m$.
If these rates are specified in angular frequencies, the relaxation time $T_{1,m}$ of mode $m$ is given by $T_{1,m} = 1/\kappa_m$.
A standard method to include the loss rates in a mathematical description of the circuit is through the Lindblad equation~\cite{johansson2012qutip}, where the losses would be included as collapse operators $\sqrt{\kappa_m}\hat a_m$

The frequencies, dissipation rates, and Kerr parameters can all be obtained via methods of the \inline{Qcircuit} object.
These methods will return numerical values, and we should always specify the values of symbolically defined circuit parameters as keyword arguments.
Lists, or Numpy arrays, can be provided here making it easy to perform parameter sweeps.
Additionally, initializing the circuit is the most computationally expensive operation, so this will be by far the fastest method to perform parameter sweeps.
We will assume that we want to determine the parameters of the Hamiltonian~(\ref{eq:hamiltonian_first_order_maintext}) for the circuit of Fig.~\ref{fig:1} at different values of $L_j$.
The values for $L_j$ are stored as a Numpy array
\begin{lstlisting}
Lj_list = numpy.linspace(11e-9,9e-9,101)
\end{lstlisting}
We can assign the frequency, dissipation rates, self-Kerr, and cross-Kerr parameters to the variables \inline{f}, \inline{k}, \inline{A} and \inline{chi} respectively, by calling
\begin{lstlisting}
f = circuit.eigenfrequencies(Lj = Lj_list)
k = circuit.loss_rates(Lj = Lj_list)
A = circuit.anharmonicities(Lj = Lj_list)
chi = circuit.kerr(Lj = Lj_list)
\end{lstlisting}
or alternatively through a single function call:
\begin{lstlisting}
f,k,A,chi = circuit.f_k_A_chi(Lj = Lj_list)
\end{lstlisting}
All values returned by these methods are given in Hertz, not in angular frequency.
With respect to the conventional way of writing the Hamiltonian, which we have also adopted in~(\ref{eq:hamiltonian_first_order_maintext}), we thus return the frequencies as $\omega_m/2\pi$, the loss rates as $\kappa_m/2\pi$ and the Kerr parameters as $A_m/h$ and $\chi_{mn}/h$.
Note that \inline{f}, \inline{k}, \inline{A}, are arrays, where the index \inline{m} corresponds to mode $m$, and modes are ordered with increasing frequencies.
For example, \inline{f[0]} will be an array of length 101, which stores the frequencies of the lowest frequency mode as \inline{Lj} is swept from 11 to 9 nH.
The variable \inline{chi} has an extra dimension, such that \inline{chi[m,n]} corresponds to the cross-Kerr between modes \inline{m} and \inline{n}, and \inline{chi[m,m]} is the self-Kerr of mode \inline{m}, which has the same value as \inline{A[m]}.
These generated values are plotted in Fig.~\ref{fig:2}.
We can also print these parameters in a visually pleasing way to get an overview of the circuit characteristics for a given set of circuit parameters.
For a Josephson inductance of $9$ nH, this is done through the command
\begin{lstlisting}
circuit.f_k_A_chi(Lj = 10e-9, pretty_print = True)
\end{lstlisting}
which will print 
\begin{lstlisting}
mode |   freq.  |   diss.  |   anha.  |
   0 | 4.99 GHz | 9.56 kHz | 10.5 kHz |
   1 | 5.28 GHz |  94.3 Hz |  189 MHz |

Kerr coefficients 
diagonal = Kerr
off-diagonal = cross-Kerr
mode |     0    |    1    |
   0 | 10.5 kHz |         |
   1 | 2.82 MHz | 189 MHz |
\end{lstlisting}
We see that mode 1 is significantly more anharmonic than mode 0, whereas mode 0 has however a higher dissipation.
We would expect that mode 1 is thus the resonance which has current fluctuations mostly located in the junction, 
whilst mode 0 is located on the other side to the coupling capacitor, where it can couple more strongly to the resistor.

\begin{figure}[ht!]
\includegraphics[width=0.5\textwidth]{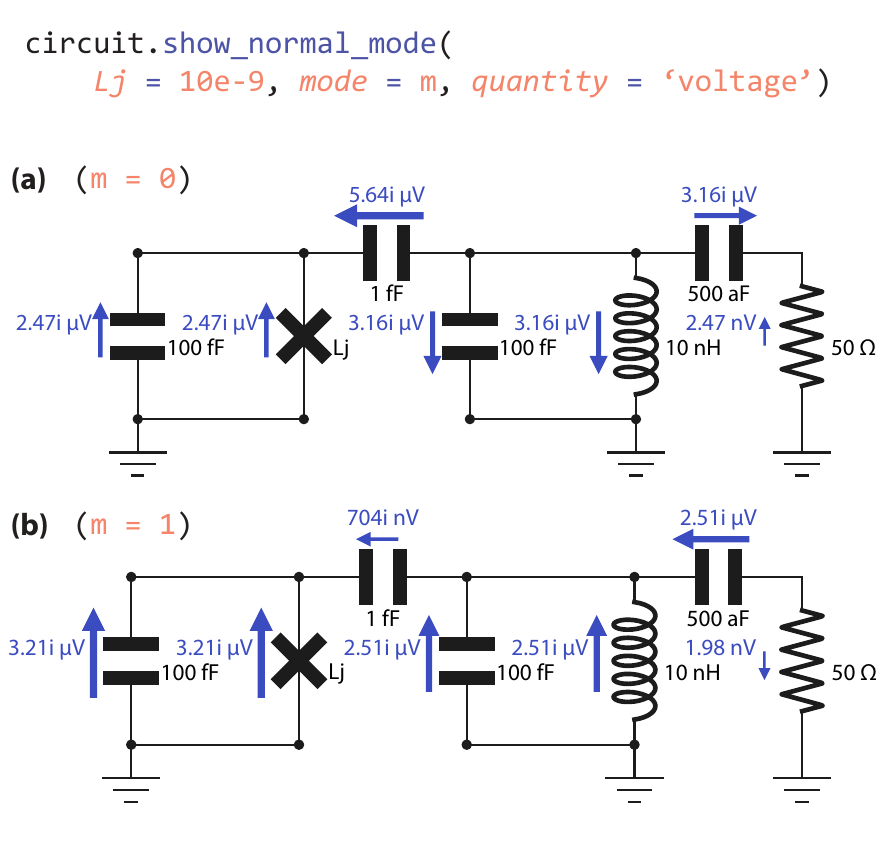}
\caption{
\textbf{Visualizing normal modes.}
The \inline{show_normal_mode} method overlays the circuit with arrows representing the voltage across components when the circuit is populated with a single-photon amplitude coherent state.
The arrows are annotated with the value of the complex voltage oscillating across a component, where the direction of the arrow indicates the direction of a phase 0 for that component.
The absolute value of this annotation corresponds to the zero-point fluctuations of the given quantity across the component.
The length and thickness of the arrows scale with the magnitude of the voltage.
\inline{show_normal_mode} takes as argument any unspecified circuit parameter, here we specify \inline{Lj=10e-9} where the two modes undergo an avoided crossing.
We plot each mode by specifying \inline{mode = 0} or \inline{mode = 1} and see that for mode \inline{0}, the anti-symmetric mode, the voltage has opposite signs on each side of the coupling capacitor, leading to a larger voltage across the coupler (and hence a larger effective capacitance and lower frequency) than the symmetric mode.
}
\label{fig:3}
\end{figure}

Such interpretations can be verified by plotting a visual representation of the normal modes on top of the circuit as explained below.
This can be done by plotting either the current, voltage, charge or flux distribution, overlaid on top of the circuit schematic.
As shown in Fig.~\ref{fig:3}, this is done by adding arrows, representing one of these quantities at each circuit component and annotating it with the value of that component.
The annotation corresponds to the complex amplitude, or phasor, of a quantity across the component, if the mode was populated with a single photon amplitude coherent state.
The absolute value of this annotation corresponds to the contribution of a mode to the zero-point fluctuations of the given quantity across the component.
The direction of the arrows indicates what direction we take for 0 phase for that component.

We note that an independantly developped Julia platform also allows the calculation of normal mode frequencies and dissipation rates for circuits~\cite{ScheerBlock2018}.

\section{Outlook}

We have presented QuCAT, a Python library to automatize and speed up the design process and analysis of superconducting circuits.
By facilitating quick tests of different circuit designs, and helping develop an intuition for the physics of quantum circuits, we also hope that QuCAT will enable users to develop even more innovative circuits.

Possible extensions of the QuCAT features could include black-box impedance components to model distributed components~\cite{nigg_black-box_2012}, more precisely modeling lossy circuits~\cite{solgun2014blackbox,solgun2015multiport}, handling static offsets in flux or charge through DC sources, additional elements such as coupled inductors or superconducting quantum interference devices (SQUIDS) and different quantization methods, enabling for example quantization in the charge or flux basis. 
The latter would extend QuCAT beyond the scope of weakly-anharmonic circuits.
In terms of performance, QuCAT would benefit from delegating analytical calculations to a more efficient, compiled language, with the exciting prospect of simulating large scale circuits~\cite{2019APS..MARA42002K}. 
Note however that there is a strong limitation on the maximum Hilbert space size that one can simulate after extracting the Hamiltonian.

\appendix

\section{Applications}

\subsection{Designing a microwave filter}
\begin{figure}[ht!]
\includegraphics[width=0.5\textwidth]{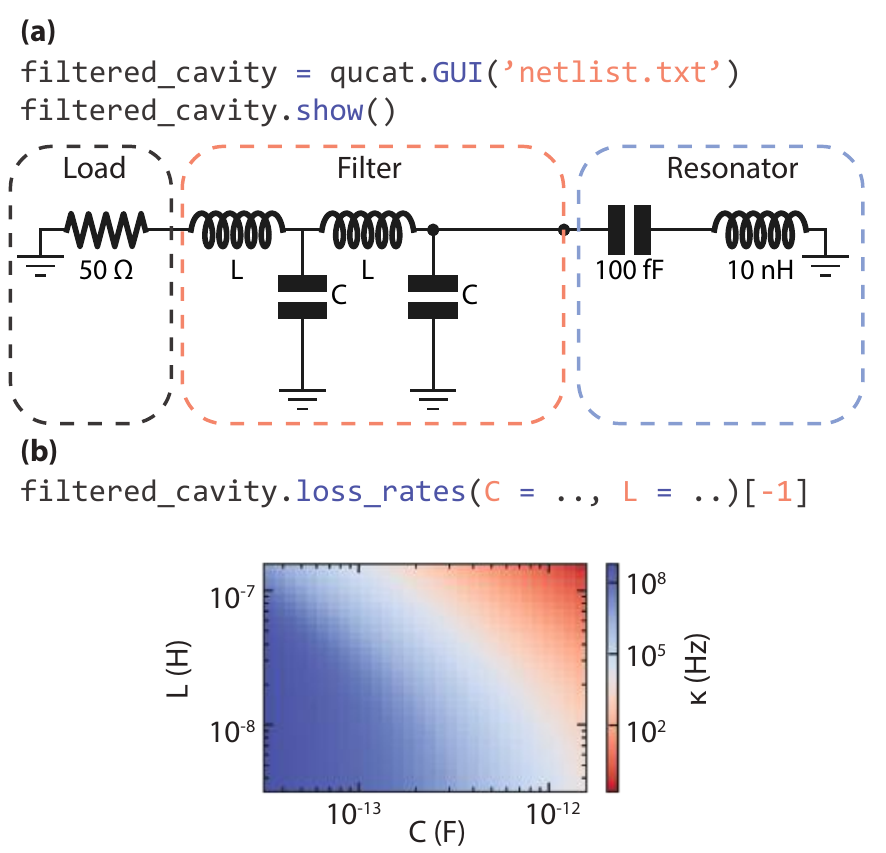}
\caption{
\textbf{Design of a microwave filter. (a) }
Using the QuCAT GUI, we build then plot a model of a filtered cavity.
A 50 $\Omega$ load, representing a cable, is connected to an LC resonator through two LC band pass filters.
\textbf{(b)} The dissipation rate of the resonator is plotted as a function of inductance and capacitance of the filter using the \inline{loss_rates} method.
}
\label{fig:filtered_cavity}
\end{figure}
In this application we show how QuCAT can be used to design classical microwave components. 
We study here a band pass filter made from two LC oscillators with the inductor inline and a capacitive shunt to ground.
Such a filter can be used to stop a DC bias line from inducing losses, whilst being galvanically connected to a resonator, see for example Ref.~\cite{Viennot2018}.
In this case we are interested in the loss rate $\kappa$ of a LC resonator connected through this filter to a 50 $\Omega$ load, which could emulate a typical microwave transmission line.
We want to study how $\kappa$ varies as a function of the inductance $L$ and capacitance $C$ of its components.

The QuCAT \inline{GUI} function can be used to open the GUI, the user will manually create the circuit, and upon closing the GUI a Qcircuit object is stored in the variable \inline{filtered_cavity}.
By calling the method \inline{show}, we display the circuit as shown in Fig.~\ref{fig:filtered_cavity}(a).
These steps are accomplished with the code
\begin{lstlisting}
# Open the GUI and manually build the circuit
filtered_cavity = qucat.GUI('netlist.txt')
# Display the circuit
filtered_cavity.show()
\end{lstlisting}
We can then access the loss rates of the different circuit modes through the method \inline{loss_rates}.
Since the values of $C$ and $L$ were not specified in the construction of the circuit, their values have to be passed as keyword arguments upon calling \inline{loss_rates}.
For example, the loss rate for a 1 pF capacitor and 100 nH inductor is obtained through
\begin{lstlisting}
# Loss rates of all modes
k_all = filtered_cavity.loss_rates(C = 1e-12, L = 100e-9)
# Resonator loss rate
k = k_all[-1]
\end{lstlisting}
Since the filter capacitance and inductance is large relative to the capacitance and inductance of the resonator, the modes associated with the filter will have a much lower frequency.
We can thus access the loss rate of the resonator by always selecting the last element of the array of loss rates with the command \inline{k_all[-1]}
The dissipation rates for different values of the capacitance and inductance are plotted in Fig.~\ref{fig:filtered_cavity}(b).
%

%

\subsection{Computing optomechanical coupling}
\begin{figure}[ht!]
\includegraphics[width=0.5\textwidth]{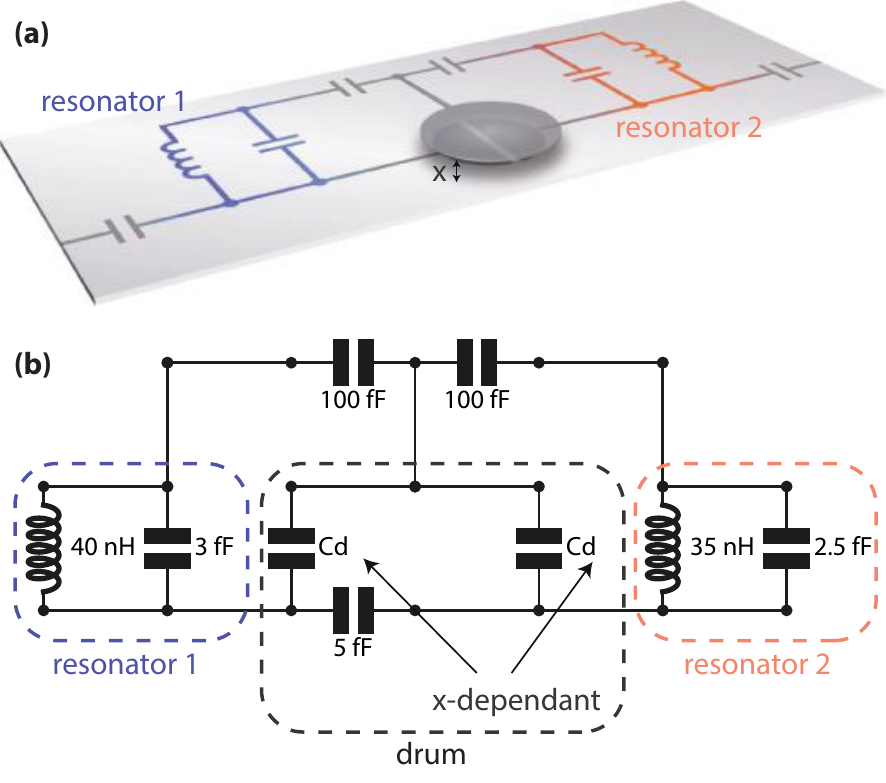}
\caption{
\textbf{Example of an optomechanical system}
\textbf{(a)} Schematic of the device, adapted from Ref.~\cite{ockeloen2016low} under a \href{https://creativecommons.org/licenses/by/3.0/}{CC BY 3.0} license. 
Two resonators are connected through a network of capacitances and a mechanically compliant capacitor (drum).
\textbf{(b)} QuCAT reconstruction of the circuit.
By specifying a label for the mechanically compliant capacitances, we have the possibility to compute the eigenfrequencies $\omega_m$ with the method \inline{eigenfrequencies} for small variation in $C_d(x)$.
This enables an easy computation of the optomechanical coupling $\propto d\omega_m/dx$.
}
\label{fig:optomechanics}
\end{figure}
In this application, we show how QuCAT can be used for analyzing another classical system, that of microwave optomechanics.
One common implementation of microwave optomechanics involves a mechanically compliant capacitor, or drum, embedded in one or many microwave resonators~\cite{Teufel2011}.
One quantity of interest is the single-photon optomechanical coupling.
This quantity is the change in mode frequency $\omega_m$ that occurs for a displacement $x_\text{zpf}$ of the drum (the zero-point fluctuations in displacement)
\begin{equation}
    g_0 = x_\text{zpf}\frac{\partial \omega_m}{\partial x}
\end{equation}
The change in mode frequency as the drum head moves $\partial \omega_m/\partial x$ is not straightforward to compute for complicated circuits.
One such example is that of Ref.~\cite{ockeloen2016low}, where two microwave resonators are coupled to a drum via a network of capacitances as shown in Fig.~\ref{fig:optomechanics}(a).
Here, we will use QuCAT to calculate the optomechanical coupling of the drums to both resonator modes of the circuit.
We start by reproducing the circuit of Fig.~\ref{fig:optomechanics}(a), excluding the capacitive connections on the far left and right.
This is done via the graphical user interface opened with the \inline{qucat.GUI} function.
Upon closing the graphical user interface, the resulting Qcircuit is stored in the variable \inline{OM}, and the \inline{show} method is used to display the schematic of Fig.~\ref{fig:optomechanics}(a).
These steps are accomplished with the code below
\begin{lstlisting}
# Open the GUI and manually build the circuit
OM = qucat.GUI('netlist.txt')
# Display the circuit
OM.show()
\end{lstlisting}
We use realistic values for the circuit components without trying to be faithful to Ref.~\cite{ockeloen2016low}, the aim of this section is to illustrate a method to obtain $g_0$.
Crucially, the mechanically compliant capacitors have been parametrized by the symbolic variable $C_d$.
We can now calculate the resonance frequencies of the circuit with the method \inline{eigenfrequencies} as a function of a keyword argument $C_d$.
The next step is to define an expression for $C_d$ as a function of the mechanical displacement $x$ of the drum head with respect to the immobile capacitive plate below it. 
\begin{lstlisting}
def Cd(x):
    # Radius of the drumhead
    radius = 10e-6
    # Formula for half a circular parallel plate capacitor
    return eps*pi*radius**2/x/2
\end{lstlisting}
where \inline{pi} and \inline{eps} have been set to the values of $\pi$ and the vacuum permittivity respectively.
We have divided the usual formula for parallel plate capacitance by 2 since, as shown in Fig.~\ref{fig:optomechanics}(a), the capacitive plate below the drum head is split in two electrodes.
We are now ready to compute $g_0$.
Following Ref.~\cite{Teufel2011}, we assume the rest position of the drum to be $D=50$ nm above the capacitive plate below.
And we assume the zero-point fluctuations in displacement to be $x_\text{zpf} = 4$ fm.
We start by differentiating the mode frequencies with respect to drum displacement using a finite differences formula
\begin{lstlisting}
# drum-capacitor gap
D = 50e-9
# difference quotient
h = 1e-18
# derivative of eigenfrequencies
G = (OM.eigenfrequencies(Cd = Cd(D+h))-OM.eigenfrequencies(Cd = Cd(D)))/h
\end{lstlisting}
\inline{G} is an array with values $2.3\times 10^{16}$ Hz.$\text{m}^{-1}$ and $3.6\times 10^{16}$ Hz.$\text{m}^{-1}$ corresponding to the lowest and higher frequency modes respectively.
Multiplying these values with the zero-point fluctuations 
\begin{lstlisting}
# zero-point fluctuations
x_zpf = 4e-15
g_0 = G*x_zpf
\end{lstlisting}
yields couplings of $96$ and $147$ Hz.
The lowest frequency mode thus has a $96$ Hz coupling to the drum.

If we want to know to which part of the circuit (resonator 1 or 2 in Fig.~\ref{fig:optomechanics}) this mode pertains, we can visualize it by calling
\begin{lstlisting}
OM.show_normal_mode(
    mode=0,
    quantity='current',
    Cd=Cd(D))
\end{lstlisting}
and we find that the current is majoritarily located in the inductor of resonator 1.
%
%
\subsection{Convergence in multi-mode cQED}\label{sec:mmusc}
\begin{figure*}[ht!]
\includegraphics[width=1\textwidth]{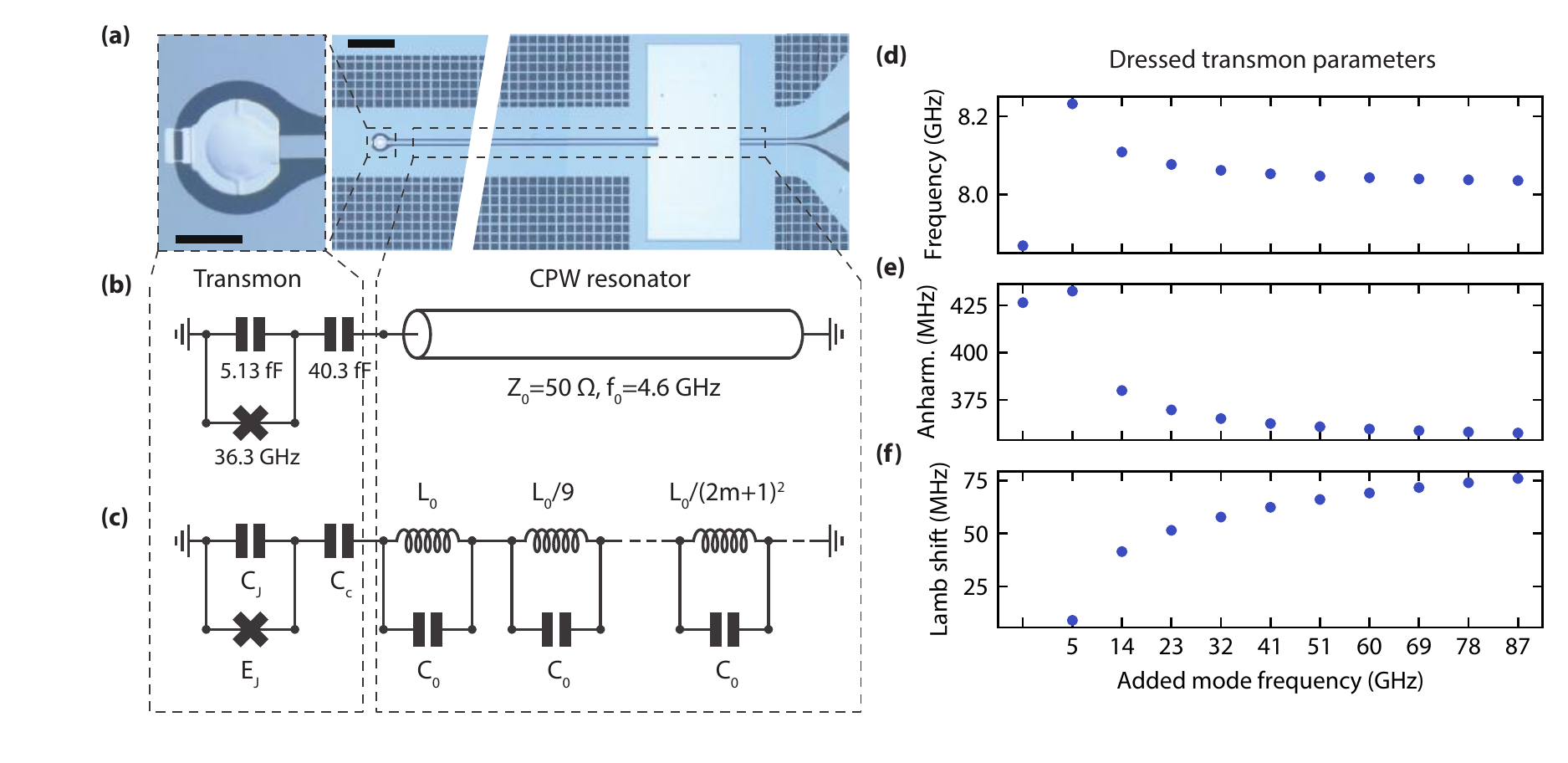}
\caption{
\textbf{Convergence in multi-mode cQED. (a) }
Optical micrograph of the device studied in this example, adapted from Ref.~\cite{bosman_multi-mode_2017} under a \href{https://creativecommons.org/licenses/by/4.0/}{CC BY 4.0} license.
Light-blue corresponds to superconductor, dark blue to an insulating substrate.
On the left we see a vacuum-gap transmon: a capacitor plate suspended over the end of a coplanar-waveguide (CPW) resonator shorted to ground through two Josephson junctions.
The scale bar corresponds to 30 $\mu$m.
On the right a CPW $\lambda/4$ resonator, capacitively coupled to the transmon on one side and shorted to ground through a large shunt capacitor on the other.
The scale bar corresponds to 100 $\mu$m.
\textbf{(b)} Circuit schematic of the device.
The CPW resonator hosts a number of modes, and is equivalent to a series assembly of LC oscillators shown in \textbf{(c)}.
This circuit is built programmatically in QuCAT, and the qubit parameters are extracted for different total numbers of modes.
In \textbf{(d)} and \textbf{(e)} we plot the transmon mode frequency $\omega_t/2\pi$ and anharmonicity $A_t/h$, where $t$ refers to the transmon-like mode, using the methods \inline{eigenfrequencies} and \inline{anharmonicities} respectively.
In \textbf{(f)} we plot the shift defined in Ref.~\cite{gely2017nature} as the Lamb shift: the shift in transmon frequency (following Eq.~\ref{eq:hamiltonian_first_order_maintext}) due solely to the vacuum-fluctuations in the other modes $\frac{1}{2}\sum_{m\ne t}\chi_{t,m}$, obtained with the \inline{kerr} method.
These calculations allow the user to gauge how many modes are relevant to the physics of the circuit.
}
\label{fig:mmusc}
\end{figure*}
In this section we use QuCAT to study the convergence of parameters in the first order Hamiltonian (Eq.~\ref{eq:hamiltonian_first_order_maintext}) of an ultra-strongly coupled multi-mode circuit QED system.
Using a length of coplanar waveguide terminated with engineered boundary conditions is a common way of building a high quality factor microwave resonator.
One implementation is a $\lambda/4$ resonator terminated on one end by a large shunt capacitor, acting as a near-perfect short circuit for microwaves such that only a small amount of radiation may enter or leave the resonator.
On the other end one places a small capacitance to ground: an open circuit.
The shunt capacitor creates a voltage node, and at the open end the voltage is free to oscillate.
This resonator hosts a number of normal modes, justifying its lumped element equivalent circuit: a series of LC oscillators with increasing resonance frequency~\cite{gely2017convergence}.
Here, we study such a resonator with a transmon circuit capacitively coupled to the open end.
In particular we consider this coupling to be strong enough for the circuit to be in the multi-mode ultra-strong coupling regime as studied experimentally in Ref.~\cite{bosman_multi-mode_2017} and theoretically in Ref.~\cite{gely2017convergence}.
The particularity of this regime is that the transmon has a considerable coupling to multiple modes of the resonator.
It then becomes unclear how many of these modes to consider for a realistic modeling of the system.
This regime is reached by maximizing the coupling capacitance of the transmon to the resonator and minimizing the capacitance of the transmon to ground.
The experimental device accomplishing this is shown in Fig.~\ref{fig:mmusc}(a), with its schematic equivalent in Fig.~\ref{fig:mmusc}(b), and the lumped-element model in Fig.~\ref{fig:mmusc}(c).

We will use QuCAT to track the evolution of different characteristics of the system as the number of considered modes $N$ increases.
For this application, programmatically building the circuit is more appropriate than using the GUI.
We start by defining some constants
\begin{lstlisting}
# fundamental mode frequency of the resonator
f0 = 4.603e9
w0 = f0*2.*numpy.pi
# characteristic impedance of the resonator
Z0 = 50
# Josephson energy (in Hertz)
Ej = 18.15e9
# Coupling capacitance
Cc = 40.3e-15
# Capacitance to ground
Cj = 5.13e-15

# Capacitance of all resonator modes
C0 = numpy.pi/4/w0/Z0
# Inductance of first resonator mode
L0 = 4*Z0/numpy.pi/w0
\end{lstlisting}
we can then generate a Qcircuit we name \inline{mmusc}, as an example here with $N=10$ modes.
\begin{lstlisting}
# Initialize list of components for Transmon and coupling capacitor
netlist = [
    qucat.J(12,1,Ej,use_E=True),
    qucat.C(12,1,Cj),
    qucat.C(1,2,Cc)]

# Add 10 oscillators
for m in range(10):
    # Nodes of m-th oscillator
    node_minus = 2+m
    node_plus = (2+m+1)
    # Inductance of m-th oscillator
    Lm = L0/(2*m+1)**2
    # Add oscillator to netlist
    netlist = netlist + [
        qucat.L(node_minus,node_plus,Lm),
        qucat.C(node_minus,node_plus,C0)]

# Create Qcircuit
mmusc = qucat.Network(netlist)
\end{lstlisting}
Note that \inline{12} is the index of the ground node.

We can now access some parameters of the system.
Only the first mode of the resonator has a lower frequency than the transmon. 
The transmon-like mode is thus indexed as mode \inline{1}.
Its frequency is given by
\begin{lstlisting}
mmusc.eigenfrequencies()[1]
\end{lstlisting}
and the anharmonicity of the transmon, computed from first order perturbation theory (see Eq.~\ref{eq:hamiltonian_first_order_maintext}) with
\begin{lstlisting}
mmusc.anharmonicities()[1]
\end{lstlisting}
Finally the Lamb shift, or shift in the transmon frequency resulting from the zero-point fluctuations of the resonator modes, is given following Eq.~(\ref{eq:hamiltonian_first_order_maintext}) by the sum of half the cross-Kerr couplings between the transmon mode and the others
\begin{lstlisting}
lamb_shift = 0
K = mmusc.kerr()
for m in range(10):
  if m!=1:
    lamb_shift = lamb_shift + K[1][m]/2
\end{lstlisting}
These parameters for different total number of modes are plotted in Figs~\ref{fig:mmusc}(d-f).
From this analysis, we find that as we reach 10, the plotted parameters are converging.
Surprisingly, adding even the highest modes significantly modifies the total Lamb shift of the Transmon despite large frequency detunings.

\subsection{Modeling a tuneable coupler}
\begin{figure*}[ht!]
\includegraphics{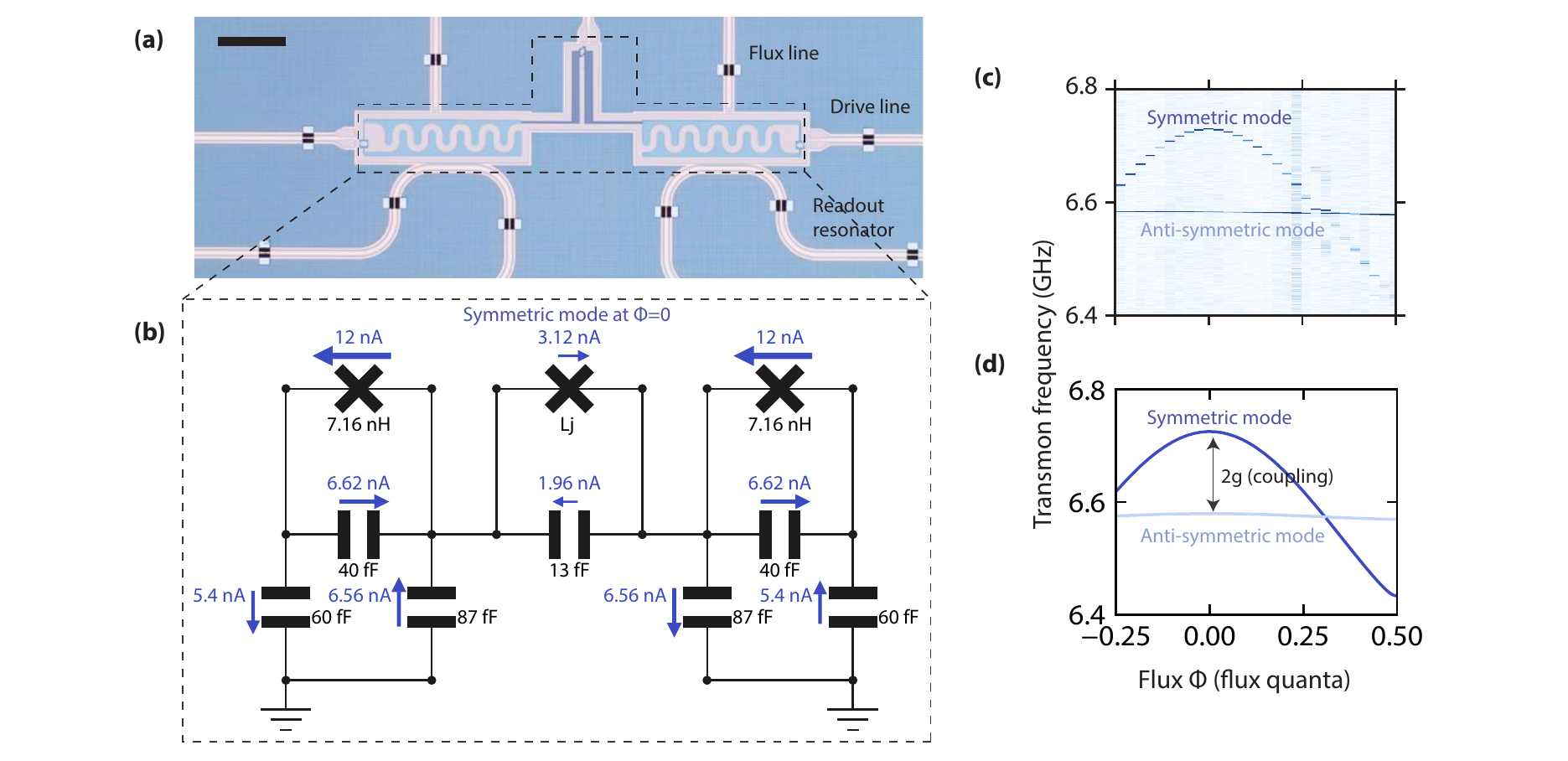}
\caption{
\textbf{Tuneable coupler circuit analysis. (a) }
Optical micrograph of the device studied in this example, adapted from Ref.~\cite{kounalakis2018tuneable} under a \href{https://creativecommons.org/licenses/by/4.0/}{CC BY 4.0} license.
We will omit the flux lines, drive lines and readout resonators for simplicity in this example, and concentrate on the part of the device in the dashed box.
The circuit consists of two near-identical transmon qubits coupled through a third ``coupler" transmon.
Scale bar corresponds to 200 $\mu$m.
\textbf{(b)} Equivalent lumped-element circuit constructed with the QuCAT GUI and displayed using the \inline{show_normal_mode} method.
This method has overlaid the circuit with the currents flowing through the components when the highest frequency mode is populated with a single-photon-amplitude coherent state.
Most of the current is located in the resonantly coupled transmons rather than the coupler, and the fact that the coupled transmons are identical leads to the symmetry on each side of the coupler.
This mode is called symmetric since the current in both coupled transmons flows in the same direction. 
The net current through the coupling junction makes the mode frequency sensitive to changes in the coupling junction inductance tuned with a superconducting quantum interference device or SQUID.
The change in symmetric mode frequency is shown in the experimental measure of the response frequencies in \textbf{(c)} (adapted from Ref.~\cite{kounalakis2018tuneable} under a \href{https://creativecommons.org/licenses/by/4.0/}{CC BY 4.0} license), and in the diagonalization of the Hamiltonian generated from QuCAT in \textbf{(d)}.
}
\label{fig:tuneable_coupler}
\end{figure*}
\begin{figure*}[ht!]
\includegraphics{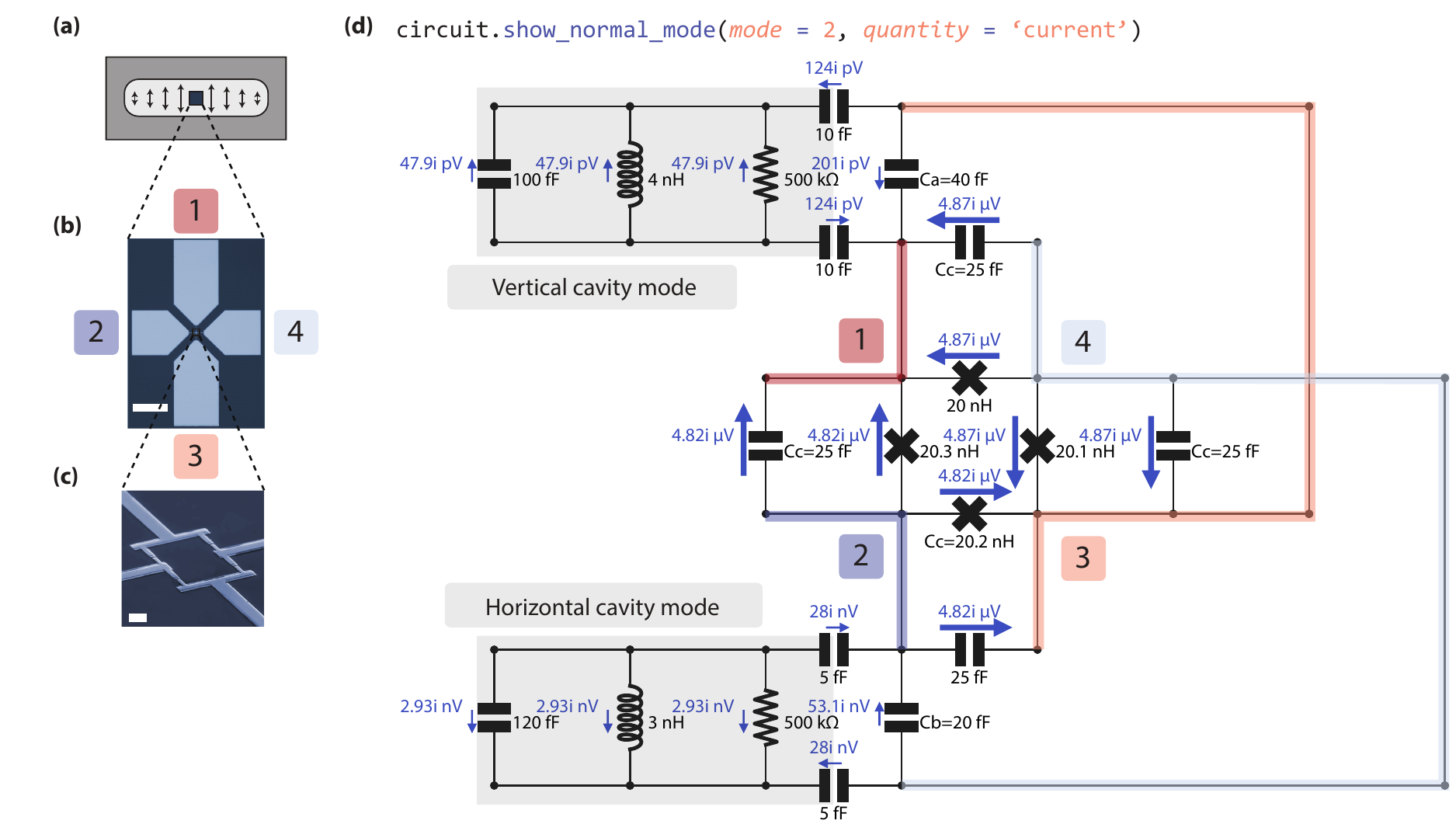}
\caption{
\textbf{Trimon device and Purcell-decay-protected mode visualization. (a) }
Schematic of the cross-cut of a 3D microwave cavity.
Dark gray shows metal whilst light gray show the hollowed out section forming the cavity.
Arrows represent the electric field of the $TE_{101}$, or ``vertical" cavity mode.
In the cavity is placed a chip hosting the trimon circuit shown in the optical micrograph \textbf{(b)}.
The circuit has 4 capacitive pads labeled from 1 to 4.
These pads are connected by the Josephson junction ring shown in the scanning electron microscope image \textbf{(c)}.
Scale bars correspond 200 and 2 $\mu$m for panels \textbf{(b)} and \textbf{(c)} respectively.
\textbf{(d)} Lumped-element equivalent circuit of the device constructed using the QuCAT GUI and displayed with \inline{show_normal_mode}.
The four pads of the trimon are color-coded to match \textbf{(b)}.
The capacitor $C_a$ formed by pads 1 and 3 forms an electrical dipole which couples to a vertical cavity mode, and the capacitor $C_b$ formed by pads 2 and 4 forms an electrical dipole which couples to modes with horizontal electric fields.
The \inline{show_normal_mode} overlays the voltage across different components if a single-photon amplitude coherent state was populating mode 2.
This mode has a particularity that the voltage is concentrated across the junctions and their parallel capacitors without leading to a buildup of voltage across the capacitors $C_a$ or $C_b$.
This decouples mode 2 from the cavity mode decay (no Purcell effect) whilst the presence of voltage fluctuations across the junctions will ensure cross-Kerr coupling to the other modes of the system.
Concerning panels (a-c): reprinted figures with permission from \href{https://doi.org/10.1103/PhysRevApplied.7.054025}{T. Roy et al., Phys. Rev. Appl. \textbf{7} (5), 054025 (2017)}. Copyright 2017 by the American Physical Society.
}
\label{fig:trimon}
\end{figure*}
\begin{figure}[ht!]
\includegraphics{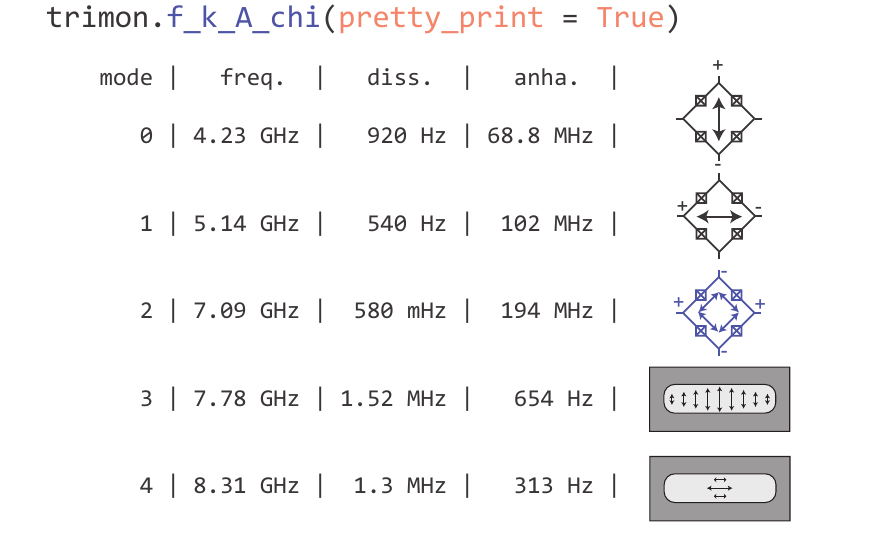}
\caption{
\textbf{Other modes of the Trimon. }
Using the \inline{f_k_A_chi} method together with the \inline{pretty_print} option gives the user an overview of the different modes frequencies, dissipations rates and levels of anharmonicity.
Here we have overlaid the output of the method with schematics of the corresponding trimon and cavity modes adapted from Ref.~\cite{roy2017implementation}.
One can identify a mode to the schematic by observing where the currents or voltages are mostly located in the circuit using the \inline{show_normal_mode} method as in Fig.~\ref{fig:trimon}(d).
Since the only resistors of the circuit are located in the cavity modes, all dissipation in transmon modes 0 through 2 are due to the Purcell effect. 
Mode 2 is better protected from this effect by 3 orders of magnitude with respect to the two other transmon modes.
Concerning schematics: reprinted figures with permission from \href{https://doi.org/10.1103/PhysRevApplied.7.054025}{T. Roy et al., Phys. Rev. Appl. \textbf{7} (5), 054025 (2017)}. Copyright 2017 by the American Physical Society.
}
\label{fig:trimon_2}
\end{figure}
In this section, we study the circuit of Ref.~\cite{kounalakis2018tuneable} where two transmon qubits are coupled through a tuneable coupler.
This tuneable coupler is built from a capacitor and a Superconducting Quantum Interference Device, or SQUID.
By flux biasing the SQUID, we change the effective Josephson energy of the coupler, which modifies the coupling between the two transmons.
We will present how the normal mode visualization tool helps in understanding the physics of the device.
Secondly, we will show how a Hamiltonian generated with QuCAT accurately reproduces experimental measurements of the device.

We start by building the device shown in Fig.~\ref{fig:tuneable_coupler}(a).
More specifically, we are interested in the part of the device in the dashed box, consisting of the two transmons and the tuneable coupler.
The other circuitry, the flux line, drive line and readout resonator could be included to determine external losses, or the dispersive coupling of the transmons to their readout resonator.
We will omit these features for simplicity here.
After opening the GUI with the \inline{qucat.GUI} function, manually constructing the circuit, then closing the GUI, the resulting \inline{Qcircuit} is stored in a variable \inline{TC}.
\begin{lstlisting}
TC = qucat.GUI('netlist.txt')
\end{lstlisting}
The inductance $L_j$ of the junction which models the SQUID is given symbolically, and will have to be specified when calling \inline{Qcircuit} functions.
Since $L_j$ is controlled through flux $\phi$ in the experiment, we define a function which translates $\phi$ (in units of the flux quantum) to $L_j$
\begin{lstlisting}
def Lj(phi):
    # maximum Josephson energy
    Ejmax = 6.5e9
    # junction asymmetry
    d = 0.0769
    # flux to Josephson energy
    Ej = Ejmax*numpy.cos(pi*phi) *numpy.sqrt(1+d**2 *numpy.tan(pi*phi)**2)
    # Josephson energy to inductance
    return (hbar/2/e)**2/(Ej*h)
\end{lstlisting}
where \inline{pi},  \inline{h}, \inline{hbar}, \inline{e} were assigned the value of $\pi$, Plancks constant, Plancks reduced constant and the electron charge respectively.

By visualizing the normal modes of the circuit, we can understand the mechanism behind the tuneable coupler.
We plot the highest frequency mode at $\phi=0$, as shown in Fig.~\ref{fig:tuneable_coupler}(b)
\begin{lstlisting}
TC.show_normal_mode(mode = 2, 
    quantity = 'current',
    Lj=Lj(0))
\end{lstlisting}
This mode is called symmetric since the currents flow in the same direction on each side of the coupler.
This leads to a net current through the coupler junction, such that the value of $L_j$ influences the oscillation frequency of the mode.
Conversely, if we plot the anti-symmetric mode instead, where currents are flowing away from the coupler in each transmon, we find a current through the coupler junction and capacitor on the order of $10^{-21}$ A.
This mode frequency should not vary as a function of $L_j$.
When the bare frequency of the coupler matches the coupled transmon frequencies, the coupler acts as a band-stop filter, and lets no current traverse.
At this point, both symmetric and anti-symmetric modes should have identical frequencies.

In Fig.~\ref{fig:tuneable_coupler}(c) this effect is shown experimentally through a measure of the first transitions of the two non-linear modes.
One is tuned with flux (symmetric mode), the other barely changes (anti-symmetric mode).
We can reproduce this experiment by generating a Hamiltonian with QuCAT and diagonalizing it with QuTiP for different values of the flux.
For example, at 0 flux, the two first two transition frequencies \inline{f1} and \inline{f2} can be generated from
\begin{lstlisting}
# generate a Hamiltonian
H = TC.hamiltonian(Lj = Lj(phi = 0), 
    excitations = [7,7], 
    taylor = 4, 
    modes = [1,2])
# diagonalize the Hamiltonian
ee = H.eigenenergies()
f1 = ee[1]-ee[0]
f2 = ee[2]-ee[0]
\end{lstlisting}
\inline{f1} and \inline{f2} is plotted in Fig.~\ref{fig:tuneable_coupler}(d) for different vales of flux and closely matches the experimental data.
Note that we have constructed a Hamiltonian with modes 1 and 2, excluding mode 0, which corresponds to oscillations of current majoritarily located in the tuneable coupler.
One can verify this fact by plotting the distribution of currents for mode 0 using the \inline{show_normal_mode} method.

This experiment can be viewed as two ``bare" transmon qubits coupled by the interaction
\begin{equation}
    \hat H_\text{int} = g\sigma_x^L\sigma_x^R
\end{equation}
where left and right transmons are labeled $L$ and $R$ and $\sigma_x$ is the $x$ Pauli operator.
The coupling strength $g$ reflects the rate at which the two transmons can exchange quanta of energy.
If the transmons are resonant a spectroscopy experiment reveals a hybridization of the two qubits, which manifests as two spectroscopic absorption peaks separated in frequency by $2g$.
From this point of view, this experiment thus implements a coupling which is tuneable from an appreciable value to near 0 coupling.

\subsection{Studying a Josephson-ring-based qubit}
In this section, we demonstrate the ability for QuCAT to analyze more complex circuits.
The experiment of Ref.~\cite{roy2017implementation} features a Josephson ring geometry, which is a Wheatstone-bridge-like circuit, typically difficult to analyze as it cannot be decomposed in series and parallel connections.
We consider the coupling of this ring to two lossy modes of a cavity, bringing the total number of modes in the circuit to 5.
We aim to understand the key feature of this circuit: that one qubit-like mode acts as a quadrupole with little coupling to the resonator modes.
The studied device consists of a 3D cavity (Fig.~\ref{fig:trimon}(a)) hosting a number of microwave modes, in which is positioned a chip patterned with the trimon circuit.
The trimon circuit has four capacitive pads in a cross shape (Fig.~\ref{fig:trimon}(b)) which have an appreciable coupling between each other making up the capacitance of the trimon qubit modes.
The two vertically (horizontally) positioned pads will couple to modes of the 3D cavity featuring vertical (horizontal) electric fields.
We will consider both a vertical and a horizontal cavity mode in our model.
We number these pads from 1 to 4 as displayed in Fig.~\ref{fig:trimon}(b).
Each pad is connected to its two nearest neighbors by a Josephson junction (Fig.~\ref{fig:trimon}(c)), forming a Josephson ring.
Using the QuCAT GUI, we build a lumped element model of this device, generating a \inline{Qcircuit} object we store in the variable \inline{trimon}.
\begin{lstlisting}
trimon = qucat.GUI('netlist.txt')
\end{lstlisting}
The cavity modes are modeled as RLC oscillators with each plate of their capacitors capacitively coupled to a pad of the trimon circuit.
The junction inductances are assigned different values, first to reflect experimental reality, but also to avoid infinities arising in the QuCAT analysis.
Indeed, the voltage transfer function of this Josephson ring between nodes 1,3 and nodes 2,4 will be exactly 0, which will cause errors when initializing the \inline{Qcircuit} object.
Component parameters are chosen to only approximatively match the experimental results of Ref.~\cite{roy2017implementation}, the objective here is to demonstrate QuCAT features rather than accurately model the experiment.
The particularity of this circuit is that it hosts a quadrupole mode.
It corresponds here to the second highest frequency mode and can be visualized by calling
\begin{lstlisting}
trimon.show_normal_mode(
    mode = 2,
    quantity = 'voltage')
\end{lstlisting}
the result of which is displayed in Fig.~\ref{fig:trimon}(d).
The voltage oscillations are majoritarily located in the junctions, indicating this is not a cavity mode, but a mode of the trimon circuit.
Crucially, the polarity of voltages across the junctions is such that the total voltage between pads 1 and 3 and the total voltage across pads 2 and 4 is 0, warranting the name of ``quadrupole mode".
Due to the orientation of the chip in the cavity, the vertically and horizontally orientated cavity modes will only be sensitive to voltage oscillations across pads 1 and 3 or 2 and 4.
This ensures that the mode displayed here is decoupled from the cavity modes, and from any loss channels they may incur.
We can verify this fact by computing the losses of the different modes, and comparing the losses of mode 2 to the other qubit-like modes of the circuit.
We perform this calculation by calling
\begin{lstlisting}
trimon.f_k_A_chi(pretty_print=True)
\end{lstlisting}
which will calculate and return the loss rates of the modes, along with their eigenfrequencies, anharmonicities and Kerr parameters.
Setting the keyword argument \inline{pretty_print} to \inline{True} prints a table containing all this information, which is shown in Fig.~\ref{fig:trimon_2}.
To be succinct, we have not shown the table providing the cross-Kerr couplings.
By using the \inline{show_normal_mode} method to plot all the other modes of the circuit, and noting where currents or voltages are majoritarily located, we can identify each mode with the schematics provided in Fig.~\ref{fig:trimon_2}.
The three lowest frequency modes are located in the trimon chip, and we notice that as expected the quadrupole mode 2 has a loss rate (due to resistive losses in the cavity modes) which is three orders of magnitude below the other two.
Despite this apparent decoupling, the quadrupole mode will still be coupled to both cavity mode through the cross-Kerr coupling, given by twice the square-root of the product of the quadrupole and cavity mode anharmonicities.

\section{Circuit quantization overview}\label{sec:circuit_quantization_overview}
\begin{figure}[ht!]
\includegraphics{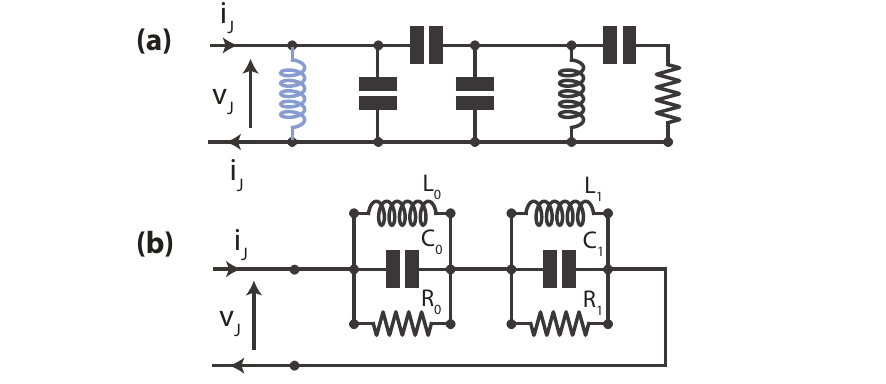}
\caption{
\textbf{Example of equivalent circuit construction to prepare for quantization.  }
We use the same example as used in Fig~\ref{fig:1}-\ref{fig:3}.
\textbf{(a)} The circuit is linearized by replacing the junction with an inductance $L_j$. The circuit is characterized at the nodes of the junction by its admittance $Y_j$.
\textbf{(b)} In the limit of small dissipation, this circuit is equivalent to a series combination of RLC resonators.
}
\label{fig:circuit_quantization}
\end{figure}
In this section we summarize the quantization method used in QuCAT, which is an expansion on the work of Ref.~\cite{nigg_black-box_2012}.
This approach is only valid in the weak anharmonic limit, where charge dispersion is negligible.
See~\cite{nigg_black-box_2012} or Sec.~\ref{sec:high_anharmonicity} for a detailed discussion of this condition.

The idea behind the quantization method is as follows.
We first consider the ``linearized" circuit.
This is a circuit where the junctions are replaced by their Josephson inductances $L_j = \phi_0^2/E_j$, where $E_j$ is the Josephson energy and the reduced flux quantum is given by $\phi_0 = \hbar/2e$.
We determine the oscillation frequencies and dissipation rates of the different normal modes of this linearized circuit.
Then, we calculate the amplitude of phase oscillations across each junction when a given mode is excited.
This will determine how non-linear each mode is.
All this information will finally allow us to build a Hamiltonian for the circuit.

\subsection{Circuit simplification to series of RLC resonators (Foster circuit)}
The eigenfrequencies and non-linearity of each mode is obtained by transforming the linearized circuit to a geometry we can easily analyze.
We will first describe this process assuming there is only a single junction in the circuit, the case of multiple junctions will follow.
We consider the example circuit of Fig.~\ref{fig:1}.
After replacing the junction with its Josephson inductance, we determine the admittance $Y_j(\omega) = I_j(\omega)/V_j(\omega)$ evaluated at the nodes of the junction.
This admittance is the inverse of the impedance measured at the nodes of the junction.
It relates the amplitude $|V_j|$ and phase $\theta(V_j)$ of the voltage oscillating at frequency $\omega$ that would build up across the junction if one would feed a current oscillating at $\omega$ with amplitude $|I_j|$ and phase $\theta(I_j)$ to one of its nodes through a infinite impedance current source.
In Fig.~\ref{fig:circuit_quantization}(a) we show a schematic describing this quantity.
In the case where all normal modes of the circuit have small dissipation rates, this circuit has an approximate equivalent shown in Fig.~\ref{fig:circuit_quantization}(b), consisting of a series of RLC resonators~\cite{solgun2014blackbox}.
By equivalent, we mean that the admittance $Y_j$ of the circuit is approximatively equal to that of a series combination of RLC resonators
\begin{equation}
  Y_j(\omega) \simeq  \frac{1}{\sum_m1/Y_m(\omega)}\
  \label{eq:Y_as_sum}
\end{equation}
which each have an admittance
\begin{equation}
  Y_m(\omega) = \frac{1}{iL_m\omega}+iC_m\omega+\frac{1}{R_m}
\end{equation}
Each RLC resonator represents a normal mode of the circuit, with resonance frequency $\omega_m = 1/\sqrt{L_mC_m}$, and dissipation rate $\kappa_m = 1/R_mC_m$.
Since this equivalent circuit comes from an extension of Foster's reactance theorem~\cite{foster1924reactance} to lossy circuits, we call this the Foster circuit.

\subsection{Hamiltonian of the Foster circuit}
The advantage of this circuit form, is that it is easy to write its corresponding Hamiltonian following standard quantization methods (see Ref.~\cite{vool2017introduction}).
In the absence of junction non-linearity, it is given by the sum of the Hamiltonians of the independent harmonic RLC oscillators:
\begin{equation}
  \sum_m \hbar \omega_m\hat a_m^\dagger \hat a_m\ .
\end{equation}
The annihilation operator $\hat a_m$ for photons in mode $m$ is related to the expression of the phase difference between the two nodes of the oscillator
\begin{equation}
\begin{split}
  \hat\varphi_{m,j} &= \varphi_{\text{zpf},m,j}(\hat a_m + \hat a^\dagger_m)\ ,\\
  \varphi_{\text{zpf},m,j} &= \frac{1}{\phi_0}\sqrt{\frac{\hbar}{2\omega_m C_m}}\ .
\label{eq:zpf}
\end{split}
\end{equation}
where $\varphi_{\text{zpf},m}$ are the zero-point fluctuations in phase of mode $m$.
The total phase difference across the Josephson junction $\hat\varphi_j$ is then the sum of these phase differences $\hat\varphi_j = \sum_m\hat\varphi_{m,j}$, and we can add the Junction non-linearity to the Hamiltonian
\begin{equation}
\hat{H} = \sum_m \hbar\omega_m\hat{a}_m^\dagger\hat{a}_m + E_j[1-\cos{\hat\varphi_j}-\frac{\hat\varphi_j^2}{2}]\ ,
\end{equation}
Since the linear part of the Hamiltonian corresponds to the circuit with junctions replaced by inductors, the linear part already contains the quadratic contribution of the junction potential $\propto\hat\varphi_j^2$, and it is subtracted from the cosine junction potential.

\subsection{Calculating Foster circuit parameters}
Both $\omega_m$ and $\kappa_m$ can be determined from $Y(\omega)$ since we have $Y(\omega_m+i\kappa_m/2)=0$ for low loss circuits.
This can be proven by noticing that the admittance $Y_m$ of mode $m$ has two zeros at
\begin{equation}
\begin{split}
  \zeta_m &= \frac{1}{\sqrt{L_mC_m}}\sqrt{1-\frac{1}{4Q^2}}+ i\frac{1}{2R_mC_m}\\
  &\simeq \omega_m + i\kappa_m/2\ .
\end{split}
\end{equation}
and $\zeta_m^*$.
The approximate equality holds in the limit of large quality factor $Q_m = R_m/\sqrt{L_m/C_m}\gg 1 $.
From Eq.~(\ref{eq:Y_as_sum}) we see that the zeros of $Y$ are exactly the zeros of the admittances $Y_k$.
The solutions of $Y(\omega)=0$, which come in conjugate pairs $\zeta_m$ and $\zeta_m^*$, thus provide us with both resonance frequencies $\omega_m = \text{Re}[\zeta_m]$ and dissipation rates $\kappa_m = 2\text{Im}[\zeta_m]$.
Additionally, we need to determine the effective capacitances $C_m$ in order to obtain the zero-point fluctuations in phase of each mode.
We focus on one mode $k$, and start by rewriting the admittance in Eq.~(\ref{eq:Y_as_sum}) as
\begin{equation}
  Y_j(\omega) =  Y_k(\omega)\frac{1}{1+\sum_{m\ne k}Y_k(\omega)/Y_m(\omega)}\ .
\end{equation}
Its derivative with respect to $\omega$ is
\begin{equation}
\begin{split}
  Y'_j(\omega) &=  Y'_k(\omega)\frac{1}{1+\sum_{m\ne k}Y_k(\omega)/Y_m(\omega)}\\
  &+Y_k(\omega)\frac{\partial}{\partial \omega}\left [\frac{1}{1+\sum_{m\ne k}Y_k(\omega)/Y_m(\omega)}\right ]\ .
\end{split}
\end{equation}
Evaluating the derivative at $\omega = \zeta_k$, where $Y_k(\zeta_k)=0$ yields
\begin{equation}
\begin{split}
  Y'_j(\zeta_k) &= Y'_k(\zeta_k) = iC_m\left(1+\frac{4Q^2}{\left(i+\sqrt{4Q^2-1}\right)^2} \right)\\
  &\simeq i2C_m\text{ for }Q_m\gg 1
\end{split}
\end{equation}
The capacitance is thus approximatively given by 
\begin{equation}
  C_m = \text{Im}\left[Y'_j(\zeta_k)\right]/2
\end{equation}

\subsection{Multiple junctions}

When more than a single junction is present, we start by choosing a single reference junction, labeled $r$.
All junctions will be again replaced by their inductances, and by using the admittance $Y_r$ across the reference junction, we can determine the Hamiltonian including the non-linearity of the reference junction through the procedure described above.
In this section, we will describe how to obtain the Hamiltonian including the non-linearity of all other junctions too
\begin{equation}
\hat H = \sum_m \hbar\omega_m\hat{a}_m^\dagger\hat{a}_m + \sum_j E_j[1-\cos{\hat\varphi_j}-\frac{\hat\varphi_j^2}{2}]\ ,
\label{eq:hamiltonian_no_taylor}
\end{equation}
where $\hat\varphi_j$ is the phase across the j-th junction.
This phase is determined by first calculating the zero-point fluctuations in phase $\varphi_{\text{zpf},m,r}$ through the reference junction $r$ for each mode $m$ given by Eq.~(\ref{eq:zpf}).
For each junction $j$, we then calculate the (complex) transfer function $T_{jr}(\omega)$ which converts phase in the reference junction to phase in junction $j$.
We can then calculate the total phase across a junction $j$ with respect to the reference phase of junction $r$, summing the contributions of all modes and both quadratures of the phase
\begin{equation}
\begin{split}
\hat\varphi_j = \sum_m\varphi_{\text{zpf},m,r}[\text{Re}\left (T_{jr}(\omega_m)\right )(\hat{a}_m+\hat{a}_m^\dagger)\\
-i\ \text{Im}\left (T_{jr}(\omega_m)\right )(\hat{a}_m-\hat{a}_m^\dagger)]
\end{split}
\label{eq:phi_j_from_phi_r}
\end{equation}

The definition of phase~\cite{vool2017introduction} $\varphi_j(t) = \phi_0^{-1}\int_{-\infty}^tv_j(\tau)d\tau$ where $v_j$ is the voltage across junction $j$ translates in the frequency domain to $\varphi_j(\omega) = i\omega\phi_0^{-1}V_j(\omega)$.
Finding the transfer function $T_{jr}$ for phase is thus equivalent to finding a transfer function for voltage $T_{jr}(\omega) = V_j(\omega)/V_r(\omega)$.
This is a standard task in microwave network analysis (see Sec.~\ref{sec:methods_network_transfer} for more details).

\subsection{Further treatment of the Hamiltonian}
The cosine potential in Eq.~(\ref{eq:hamiltonian_no_taylor}) can be expressed in the Fock basis by Taylor expanding it around small values of the phase.
This yields
\begin{equation}
\begin{split}
\hat{H} &= \sum_{m} \hbar\omega_m\hat{a}_m^\dagger\hat{a}_m\\
&+ \sum_j\sum_{n\ge 2}E_j\frac{(-1)^{n+1}}{(2n)!}\left[\sum_{m}\varphi_{\text{zpf},m,j}(\hat{a}_m^\dagger+\hat{a}_m)\right]^{2n}
\end{split}
\label{eq:hamiltonian_taylor_SI}
\end{equation}
which is the form returned by the QuCAT \inline{hamiltonian} method.
By keeping only the fourth power in the Taylor expansion and performing first order perturbation theory, we obtain 
\begin{equation}
\begin{split}
\hat{H} = \sum_m\sum_{n\ne m} (\hbar\omega_m-A_m-\frac{\chi_{mn}}{2})\hat{a}_m^\dagger\hat{a}_m \\
-\frac{A_m}{2}\hat{a}_m^\dagger\hat{a}_m^\dagger\hat{a}_m\hat{a}_m -\chi_{mn}\hat{a}_m^\dagger\hat{a}_m\hat{a}_n^\dagger\hat{a}_n
\label{eq:hamiltonian_first_order}
\end{split}
\end{equation}
Where the anharmonicity or self-Kerr of mode m is
\begin{equation}
A_m = \sum_j A_{m,j}
\end{equation}
as returned by the \inline{anharmonicites} method,
where
\begin{equation}
A_{m,j} = \frac{E_j}{2}\varphi_{\text{zpf},m,j}^4
\label{eq:Amj}
\end{equation}
is the contribution of junction j to the total anharmonicity of a mode m.
The cross-Kerr coupling between mode m and n is
\begin{equation}
\chi_{mn} = 2\sum_j \sqrt{A_{m,j}A_{n,j}}\ .
\end{equation}
Both self and cross-Kerr parameters are computed by the \inline{kerr} method.
Note in Eq.~\ref{eq:hamiltonian_first_order} that the harmonic frequency of the Hamiltonian is shifted by $A_m$ and $\sum_{n\ne m}\chi_{nm}/2$.
The former comes from the change in Josephson inductance induced by phase fluctuations of mode $m$. 
The latter is called the Lamb shift~\cite{gely2017nature} and is induced by phase fluctuations of the other modes of the circuit.

\section{Algorithmic methods}

There are three calculations to accomplish in order to obtain all the parameters necessary to write the circuit Hamiltonian. We need:
\begin{itemize}
\item the eigen-frequencies $\omega_m$ and loss rates $\kappa_m$ fulfilling $Y_r(\zeta_m=\omega_m+i\kappa_m/2) = 0$ where $Y_r$ is the admittance across a reference junction
\item the derivative of this admittance evaluated at $\zeta_m$
\item the transfer functions $T_{jr}$ between junctions $j$ and the reference junction $r$
\end{itemize}

In this section, we cover the algorithmic methods used to calculate these three quantities

\subsection{Resonance frequency and dissipation rate}\label{sec:methods_resonance_frequency}
\subsubsection{Theoretical background}

In order to obtain an expression for the admittance across the reference junction, we start by writing the set of equations governing the physics of the circuit.
We first determine a list of nodes, which are points at which circuit components connect.
Each node, labeled $n$, is assigned a voltage $v_n$.
We name $r_\pm$ the positive and negative nodes of the reference junction.

We are interested in the steady-state oscillatory behavior of the system.
We can thus move to the frequency domain, with complex node voltages $|V_n(\omega)| e^{i(\omega t+\theta(V_n(\omega_n)))}$, fully described by their phasors, the complex numbers $V_n = |V_n(\omega)| e^{i\theta(V_n(\omega_n))}$.
In this mathematical construct, the real-part of the complex voltages describes the voltage one would measure at the node in reality.
Current conservation dictates that the sum of all currents arriving at any node $n$, from the other nodes $k$ of the circuit should be equal to the oscillatory current injected at node $n$ by a hypothetical, infinite impedance current source.
This current is also characterized by a phasor $I_n$.
This can be compactly written as
\begin{equation}
    \sum_{k\ne n} Y_{nk}(V_n-V_k) = I_n
    \label{eq:kirchhoff_1}
\end{equation}
where $k$ label the other nodes of the circuit and $Y_{nk}$ is the admittance directly connecting nodes $k$ and $n$.
Note that in this notation, if a node $k_1$ can only reach node $n$ through another node $k_2$, then $Y_{nk_1} = 0$.
Inductors (with inductance $L$), capacitors (with capacitance $C$) and resistors (with resistance $R$) then have admittances $1/iL\omega$, $iC\omega$ and $1/R$ respectively.
Expanding Eq.~\ref{eq:kirchhoff_1} yields
\begin{equation}
    (\sum_{k\ne n} Y_{nk})V_n-\sum_{k\ne n} Y_{nk}V_k = I_n
\end{equation}
which can be written in matrix form as
\begin{equation}
    \begin{pmatrix}
    \Sigma_{k\ne 0}Y_{0k} &-Y_{01}  &\cdots  &-Y_{0N} \\ 
    -Y_{10} &\Sigma_{k\ne 1}Y_{1k} &\cdots  &-Y_{1N} \\ 
    \vdots &\vdots  &\ddots  & \vdots\\ 
    -Y_{N0} &-Y_{N1}  &\cdots  &\Sigma_{k\ne N}Y_{Nk} 
    \end{pmatrix}
    \begin{pmatrix}
    V_0\\ 
    V_1\\ 
    \vdots\\ 
    V_N
    \end{pmatrix}=
    \begin{pmatrix}
    I_0\\ 
    I_1\\ 
    \vdots\\ 
    I_N
    \end{pmatrix}
    \label{eq:admittance_matrix_no_ground}
\end{equation}
Since voltage is the electric potential of a node relative to another, we still have the freedom of choosing a ground node. 
Equivalently, conservation of currents imposes that current exciting that node is equal to the sum of currents entering the others, there is thus a redundant degree of freedom in Eq.(\ref{eq:admittance_matrix_no_ground}).
For simplicity, we will choose node 0 as ground.
Since we are only interested in the admittance across the reference junction, we set all currents to zero, except the currents entering the positive and negative reference junction nodes: $I_{r_+}$ and $I_{r_-} = -I_{r_+} $  respectively.
The admittance is defined by $Y_r = I_{r_+}/(V_{r_+}-V_{r_-})$.
The equations then reduce to
\begin{equation}
    \mathbf{Y}\begin{pmatrix}
    V_1\\ 
    V_2\\ 
    \vdots\\ 
    V_N
    \end{pmatrix}=Y_r
    \begin{pmatrix}
    \vdots\\ 
    0\\ 
    V_{r_+}-V_{r_-}\\ 
    0\\
    \vdots\\ 
    0\\ 
    V_{r_-}-V_{r_+}\\ 
    0\\
    \vdots
    \end{pmatrix}
    \label{eq:admittance_matrix_equation}
\end{equation}
Where $\mathbf{Y}$ is the admittance matrix
\begin{equation}
  \mathbf{Y} = 
    \begin{pmatrix}
    \Sigma_{k\ne 1}Y_{1k} &-Y_{12}  &\cdots  &-Y_{1N} \\ 
    -Y_{21} &\Sigma_{k\ne 1}Y_{2k} &\cdots  &-Y_{2N} \\ 
    \vdots &\vdots  &\ddots  & \vdots\\ 
    -Y_{N1} &-Y_{N2}  &\cdots  &\Sigma_{k\ne N}Y_{Nk} 
    \end{pmatrix}
    \label{eq:admittance_matrix}
\end{equation}

For $Y_r = 0$, Eq~\ref{eq:admittance_matrix_equation} has a solution for only specific values of $\omega = \zeta_m$.
These are the values which make the admittance matrix singular, i.e. which make its determinant zero
\begin{equation}
    \text{Det}\left[\mathbf{Y}(\zeta_m)\right] = 0
    \label{eq:det}
\end{equation}
The determinant is a polynomial in $\omega$, so the problem of finding $\zeta_m = \omega_m+i\kappa_m/2$ reduces to finding the roots of this polynomial.
Note that plugging $\zeta_m$ into the frequency domain expression for the node voltages yields $V_k(\zeta_m)e^{i\omega_mt}e^{-\kappa_mt/2}$, such that the energy $\propto v_k(t)^*v_k(t) \propto e^{-\kappa_mt}$ decays at a rate $\kappa_m$, which explains the division by two in the expression of $\zeta_m$.
Also note, that we would have obtained equation Eq.~(\ref{eq:det}) regardless of the choice of reference element.

\subsubsection{Algorithm}
\begin{figure}[ht!]
\includegraphics{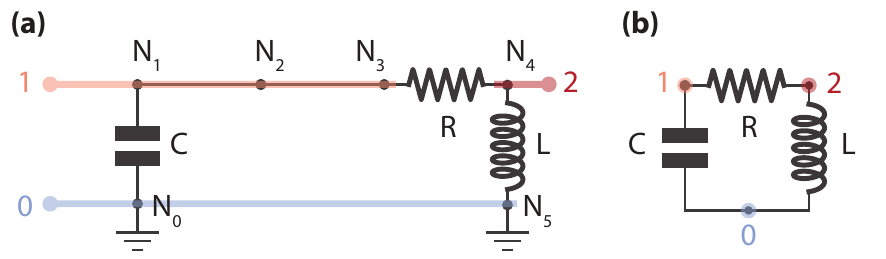}
\caption{
\textbf{Example circuit to illustrate the mode frequency finder algorithm. }
\textbf{(a)} Example of a circuit built through the GUI by a user.
\textbf{(b)} Application of the first step of the algorithm which removes the wires and grounds to obtain a minimal number of nodes without removing any components.
}
\label{fig:f_finder_example}
\end{figure}

We now describe the algorithm used to determine the solutions $\zeta_m = \omega_m+i\kappa_m/2$ of Eq.~\ref{eq:det}.
As an example, we consider the circuit of Fig.~\ref{fig:f_finder_example}(a) that a user would have built with the GUI.
The algorithm is as follows
\begin{enumerate}
    \item  Eliminate wires and grounds. In this case, nodes $N_0,N_5$ would be grouped under a single node labeled 0 and nodes $N_1,N_2,N_3$ would be grouped under node 1, we label node $N_4$ node 2, as shown in Fig.~\ref{fig:f_finder_example}(b).
    \item Compute the un-grounded admittance matrix. For each component present between the different couples of nodes, we append the admittance matrix with the components admittance. The matrix is then multiplied by $\omega$ such that all components are polynomials in $\omega$, ensuring that the determinant is also a polynomial. In this example, the matrix is
    \begin{equation}
        \begin{pmatrix}
        iC\omega^2+1/iL & -iC\omega^2  & -1/iL  \\
        -iC\omega^2 &iC\omega^2+\omega/R  &-\omega/R  \\
         -1/iL &-\omega/R  &1/iL +\omega/R
        \end{pmatrix}
    \end{equation}
    \item Choose a ground node. The node which has a corresponding column with the most components is chosen as the ground node (to reduce computation time). These rows and columns are erased from the matrix, yielding the final form of the admittance matrix
    \begin{equation}
    \mathbf{Y}=
        \begin{pmatrix}
        iC\omega^2+\omega/R  &-\omega/R  \\
        -\omega/R  &1/iL +\omega/R
        \end{pmatrix}
    \end{equation}
    \item Compute the determinant. Even if the capacitance, inductance and resistance were specified numerically, the admittance matrix would still be a function of the symbolic variable $\omega$. We thus rely on a symbolic Berkowitz determinant calculation algorithm~\cite{berkowitz1984computing,kerber2009division} implemented in the Sympy library through the \inline{berkowitz_det} function. In this example, one would obtain
    \begin{equation}
        \text{Det}[\mathbf{Y}] = LC\omega^2-iRC\omega -1\ .
    \end{equation}
    \item Find the roots of the polynomial. Whilst the above steps have to be performed only once for a given circuit, this one should be performed each time the user edits the value of a component. The root-finding is divided in the following steps as prescribed by Ref.~\cite{press2007numerical}.
    \subitem Diagonalize the polynomials companion matrix~\cite{horn1985cr} to obtain an exhaustive list of all roots of the polynomial. This is implemented in the NumPy library through the \inline{roots} function.
    \subitem Refine the precision of the roots using multiple iterations of Halley's gradient based root finder~\cite{press2007numerical} until iterations do not improve the root value beyond a predefined tolerance given by the \inline{Qcircuit} argument \inline{root_relative_tolerance}. The maximum number of iterations that may be carried out is determined by the \inline{Qcircuit} argument \inline{root_max_iterations}. If the imaginary part relative to the real part of the root is lower than the relative tolerance, the imaginary part will be set to zero. The relative tolerance thus sets the highest quality factor that QuCAT can detect.
    \subitem Remove identical roots (equal up to the relative tolerance), roots with negative imaginary or real parts, 0-frequency roots, roots for which $Y_l'(\omega_m)<0$ for all $l$, where $Y_l$ is the admittance evaluated at the nodes of an inductive element $l$, and roots for which $Q_m<$\inline{Qcircuit.Q_min}. The user is warned of a root being discarded when one of these cases is unexpected.
\end{enumerate}

The roots $\zeta_m$ obtained through this algorithm are accessed through the method \inline{eigenfrequencies} which returns the oscillatory frequency in Hertz of all the modes ${\text{Re}[\zeta_m]/2\pi}$ or \inline{loss_rates} which returns ${2\text{Im}[\zeta_m]/2\pi}$.

\subsection{Derivative of the admittance}\label{sec:methods_network_dY}
The zero-point fluctuations in phase $\varphi_{\text{zpf},m,r}$ for each mode $m$ across a reference junction $r$ is the starting point to computing a Hamiltonian for the non-linear potential of the Junctions.
As expressed in Eq.~(\ref{eq:zpf}), this quantity depends on the derivative $Y_r'$ of the admittance $Y_r$ calculated at the nodes of the reference element.
In this section we first cover the algorithm used to obtain the admittance at the nodes of an arbitrary component.
From this admittance we then describe the method to obtain the derivative of the admittance on which $\varphi_{\text{zpf},m,r}$ depends
Finally we describe how to choose a (mode-dependent) reference element.

\subsubsection{Computing the admittance}
\begin{figure}[ht!]
\includegraphics[width=0.5\textwidth]{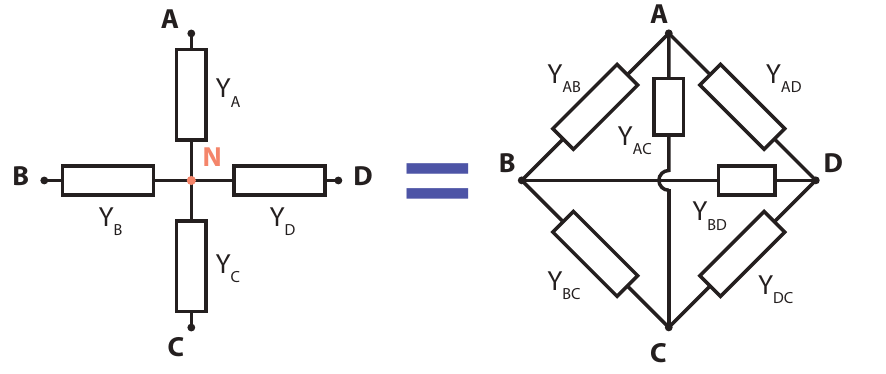}
\caption{
\textbf{Star-mesh transform.}
A node $N$ connected to nodes $A,B,C,..$ through admittances $Y_A,Y_B,...$ can be eliminated if we interconnect nodes $A,B,C,..$ with impedances $Y_{AB},Y_{AC},Y_{BC},...$ given by $Y_{XY} = Y_XY_Y/\sum_MY_M$.
We show the 5 node case, the initial network on the left is called the ``star", which is then transformed to the ``mesh" on the right, reducing the total number of nodes by 1.
}
\label{fig:star_mesh}
\end{figure}

\begin{figure}[ht!]
\includegraphics{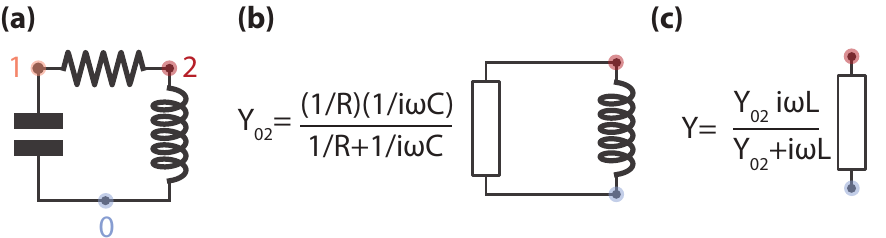}
\caption{
\textbf{Example to illustrate the admittance calculation algorithm. }
\textbf{(a)} Example of a circuit built through the GUI by a user, after removal of all wires and grounds.
\textbf{(b)} Application of the star-mesh transformation to remove node 1.
\textbf{(c)} After each application of the star-mesh transformation, parallel connections are grouped together. Only the two nodes across which we want to compute the admittance remain, the admittance is that of the remaining ``admittance component".
}
\label{fig:Y_computation_example}
\end{figure}

Here we describe a method to compute the admittance of a network between two arbitrary nodes.
We will continue using the example circuit of Fig.~\ref{fig:f_finder_example}, assuming we want to compute the admittance at the nodes of the inductor.
\begin{enumerate}
    \item Eliminate wires and grounds as in the resonance finding algorithm, nodes $N_0,N_5$ would be grouped under a single node labeled 0 and nodes $N_1,N_2,N_3$ would be grouped under node 1, we label node $N_4$ node 2. We thus obtain Fig.~\ref{fig:Y_computation_example}(a)
    \item Group parallel connections. Group all components connected in parallel as a single ``admittance component" equal to the sum of admittances of its parts. In this way two nodes are either disconnected, connected by a single inductor, capacitor, junction or resistor, or connected by a single ``admittance component".
    \item Reduce the network through star-mesh transformations. Excluding the nodes across which we want to evaluate the admittance, we utilize the star-mesh transformation described in Fig.~\ref{fig:star_mesh} to reduce the number of nodes in the network to two. If following a star-mesh transformation, two components are found in parallel, they are grouped under a single ``admittance component" as described previously. For a node connected to more than 3 other nodes the star-mesh transformation will increase the total number of components in the circuit. So we start with the least-connected nodes to maintain the total number of components in the network to a minimum. In this example, we want to keep nodes 0 and 2, but remove node 1, a start-mesh transform leads to the circuit of Fig.~\ref{fig:Y_computation_example}(b) then grouping parallel componnets leads to (c).
    \item The admittance is that of the remaining ``admittance component" once the network has been completely reduced to two nodes.
\end{enumerate}
The symbolic variables at this stage (Sympy \inline{Symbols}) are $\omega$, and the variables corresponding to any component with un-specified values.


\subsubsection{Differentiating the admittance}
The expression for the admittance obtained from the above algorithm will necessarily be in the form of multiple multiplication, divisions or additions of the admittance of capacitors, inductors or resistors.
It is thus possible to transform $Y$ to a rational function of $\omega$
\begin{equation}
    Y(\omega) = \frac{P(\omega)}{Q(\omega)} = \frac{p_0+p_1\omega+p_2\omega^2+...}{q_0+q_1\omega+q_2\omega^2+...}
\end{equation}
with the sympy function \inline{together}.
It is then easy to symbolically determine the derivative of Y, ready to be evaluated at $\zeta_m$ once the coefficients $p_i$ and $q_i$ have been extracted
\begin{equation}
\begin{split}
    Y'(\zeta_m) &= \left(P'(\zeta_m)Q(\zeta_m) -P(\zeta_m)Q'(\zeta_m)\right)/Q(\zeta_m)^2\\
      &=(p_1+2p_2\zeta_m+...)/(q_0+q_1\zeta_m+...)\ ,
\end{split}
\end{equation}
taking advantage of the property $P(\zeta_m)\propto Y(\zeta_m)=0$.

\subsubsection{Choice of reference element}
\begin{figure}[ht!]
\includegraphics{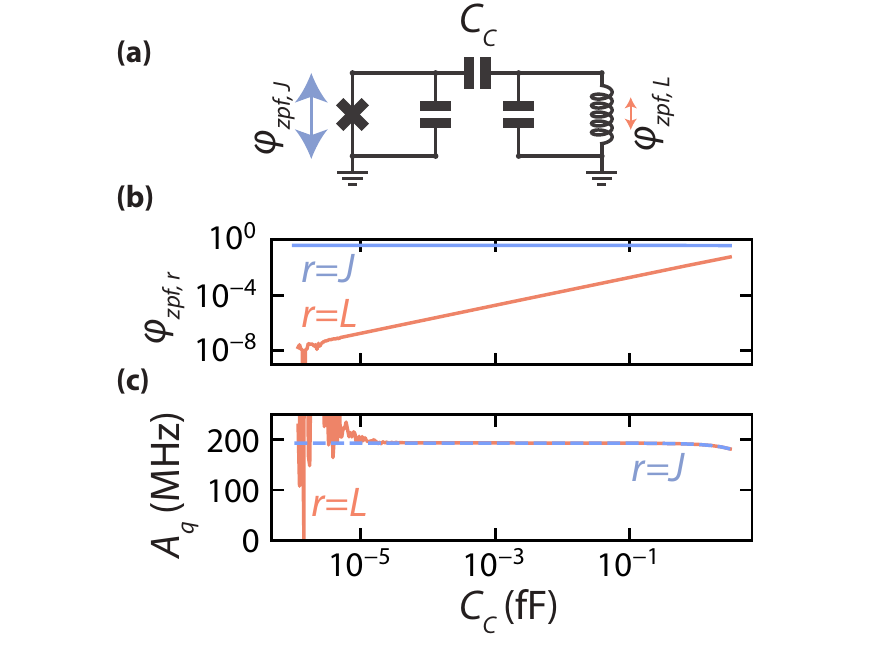}
\caption{
\textbf{Impact of the choice of reference element. }
\textbf{(a)} Schematic of the circuit used in this example.
We have used a 100 fF capacitances, a 10 nH inductor and a 8 nH Josephson inductance, we will vary the coupling capacitance.
The zero-point fluctuations in phase $\varphi_{\text{zpf},r}$ across the inductor ($r=L$) and junction ($r=J$) for most anharmonic mode are drawn on the schematic and plotted in \textbf{(b)} for different values of the coupling capacitor $C_C$.
The phase oscillations associated with this mode are mostly located in the junction, so as the coupling capacitor is lowered, the amplitude of phase oscillations diminishes in the inductor.
Below $C_C\sim 10^{-5}$ fF, numerical accuracies lead to unreliable values of the phase fluctuations in the inductor.
This results in the anharmonicity of the qubit-like mode $A_q$, plotted in \textbf{(c)}, to be incorrectly estimated if the inductor is chosen as a reference element and the anharmonicity is computed using Eq.~(\ref{eq:phi_j_from_phi_r}).
}
\label{fig:best_ref_elt}
\end{figure}

For each mode $m$, we use as reference element $r$ the inductor or junction which maximizes $\varphi_{\text{zpf},m,r}$ as specified by Eq.~\ref{eq:zpf}.
This corresponds to the element where the phase fluctuations are majoritarily located.
We find that doing so considerably increases the success of evaluating $Y'(\omega_m)$.
As an example, we plot in Fig.~\ref{fig:best_ref_elt} the zero-point fluctuations in phase $\varphi_{\text{zpf},m,r}$ of the transmon-like mode, calculated for the circuit of Fig.~\ref{fig:1}, with the junction or inductor as reference element.
What we find is that if the coupling capacitor becomes too small, resulting in modes which are nearly totally localized in either inductor or junction, choosing the wrong reference element combined with numerical inaccuracies leads to unreliable values of $\varphi_{\text{zpf},m,r}$.

\subsection{Transfer functions}\label{sec:methods_network_transfer}

\begin{figure}[ht!]
\includegraphics{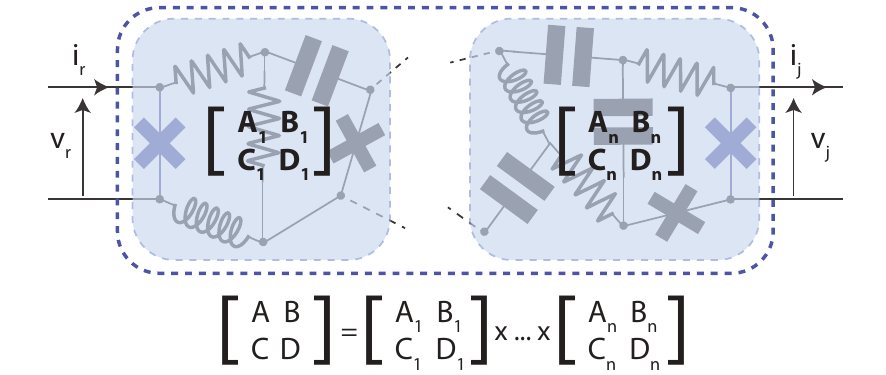}
\caption{
\textbf{Visual summary of the notations and properties of the ABCD matrix applied to the calculation of $T_{rj}$. }
The transfer function $T_{rj} = 1/A$ is the inverse of the first coefficient of the ABCD matrix which relates the voltages and currents on either end of a network.
These currents and voltages are defined as shown above, with the reference junction on the left and junction $j$ on the right.
Currents are defined as entering and exciting on the left and right respectively.
If the circuit is constituted of a cascade of two port sub-networks, the product of the sub-network ABCD matrices are equal to the ABCD matrix of the total network.
}
\label{fig:two_port_network}
\end{figure}

\begin{figure}[ht!]
\includegraphics[width=0.5\textwidth]{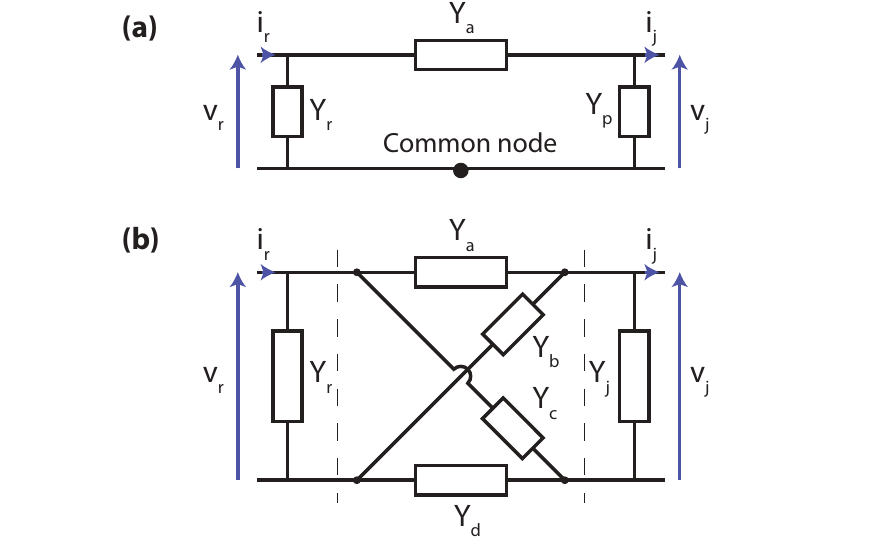}
\caption{
\textbf{Networks after star-mesh reduction.}
The two non-trivial situations reached after applying star-mesh transformations to a network to obtain $T_{rj}$.
}
\label{fig:network}
\end{figure}

In this section, we describe the method used to determine the transfer function $T_{jr}$ between a junction $j$ and the reference junction $r$.
This quantity can be computed from the ABCD matrix~\cite{pozar2009microwave}.
The ABCD matrix relates the voltages and cu{}rrents in a two port network
\begin{equation}
    \begin{pmatrix}
    V_r \\
    I_r
    \end{pmatrix}=
    \begin{pmatrix}
    A&B \\
    C&D
    \end{pmatrix}
    \begin{pmatrix}
    V_j\\
    I_j\\
    \end{pmatrix}
\end{equation}
where the convention for current direction is described in Fig.~\ref{fig:two_port_network}.
By constructing the network as in Fig.~\ref{fig:two_port_network}, with the reference junction on the left and junction $j$ on the right, the transfer function is given by
\begin{equation}
 T_{jr}(\omega) = \frac{V_j(\omega)}{V_r(\omega)} = \frac{1}{A}
\end{equation}
To determine $A$, we first reduce the circuit using star-mesh transformations (see Fig.\ref{fig:star_mesh}), and group parallel connections as described in the previous section, until only the nodes of junctions $r$ and $j$ are left.
If the junctions initially shared a node, the resulting circuit will be equivalent to the network shown in Fig.\ref{fig:network} (a).
In this case, 
\begin{equation}
    A = \left(1+\frac{Y_p}{Y_a}\right)\ .
\end{equation}
%
%
If the junctions do not share nodes, the resulting circuit will be equivalent to the network shown in Fig.\ref{fig:network}(b), where some admittances may be equal to $0$ to represent open circuits.
To compute the ABCD matrix of this resulting circuit, we make use of the property illustrated in Fig.~\ref{fig:two_port_network}: the ABCD matrix of a cascade connection of two-port networks is equal to the product of the ABCD matrices of the individual networks.
We first determine the $ABCD$ matrix of three parts of the network (separated by dashed line in Fig.\ref{fig:network}) such that the $ABCD$ matrix of the total network reads
\begin{equation}
\begin{bmatrix}
    A & B   \\
    C & D  \\
\end{bmatrix} = 
\begin{bmatrix}
    1 & 0   \\
    Y_r & 1  \\
\end{bmatrix}
\begin{bmatrix}
    \tilde A  & \tilde B   \\
    \tilde C & \tilde D  \\
\end{bmatrix}
\begin{bmatrix}
    1 & 0   \\
    Y_j & 1  \\
\end{bmatrix}\ .
\end{equation}
\begin{equation}
  A = \tilde A+\tilde B Y_j\ ,
\end{equation}
where the $A$ and $B$ coefficients of the middle part of the network are
\begin{equation}
\begin{split}
    \tilde A&=(Y_a + Y_b)(Y_c + Y_d)/(Y_a Y_d-Y_b Y_c)\\
    \tilde B&=(Y_a + Y_b + Y_c + Y_d)/(Y_a Y_d-Y_b Y_c)\ .
\end{split}
\end{equation}
The ABCD matrix for the middle part of the circuit is derived in Sec. 10.11 of Ref.~\cite{arshad2010network}, and the ABCD matrices for the circuits on either sides are provided in Ref.~\cite{pozar2009microwave}.

This method is also applied to calculate the transfer function to capacitors, inductors and resistors, notably to visualize the normal mode with the \inline{show_normal_mode} function.

\subsection{Alternative algorithmic methods}

Since symbolic calculations are the most computationally expensive steps in a typical use of QuCAT, we cover in this section some alternatives to the methods previously described, and the reasons why they were not chosen.
\subsubsection{Eigen-frequencies from the zeros of admittance}
One could solve $Y_r(\omega)=0$ where $Y_r$ is the admittance computed as explained in Sec.~\ref{sec:methods_network_dY}.
Providing good initial guesses for all values of the zeros $\zeta_m$ can be provided, a number of root-finding algorithms can then be used to obtain final values of $\zeta_m$.
A set of initial guesses could be obtained by noticing that $Y_r$ is a rational function of $\omega$.
Roots of its numerator are potentially zeros of Y, and a complete set of them is easy to obtain through a diagonalization of the companion matrix as discussed before.
Note that if these roots are roots of the denominator with equal or higher multiplicity, then they are not zeros of Y.
They can, however, make good initial guesses of a root-finding algorithm run on $Y_r$.
This requires a simplification of $Y_r$, as computed through star-mesh transforms, to its rational function form.
We find this last step to be as computationally expensive as obtaining a determinant.

A different approach, which does not require using a root-finding algorithm on $Y_r$, is to simplify the rational-function form of Y such that the numerator and denominator share no roots.
This can be done by using the extended Euclidian algorithm to find the greatest common polynomial divisor (GCD) of the numerator and denominator.
However, the numerical inaccuracies in the numerator and denominator coefficients may make this method unreliable.
The success of both of these approaches is dependent on determining a good reference component $r$, which may be mode-dependent (see Fig.~\ref{fig:best_ref_elt}).
This reference component is difficult to pick at this stage, when the mode frequencies are unknown.

\subsubsection{Finite difference estimation of the admittance derivative}
Rather than symbolically differentiating the admittance, one could use a numerical finite difference approximation, for example
\begin{equation}
     Y_r'(\omega)\simeq \frac{Y_r(\omega+\delta\omega/2)-Y_r(\omega-\delta\omega/2)}{\delta\omega}\ .
     \label{eq:finite_difference_dY}
 \end{equation} 
$Y_r$ can be obtained through star-mesh reductions, or from a resolution of Eq.~\ref{eq:admittance_matrix_equation}.

\begin{figure}[ht!]
\includegraphics[width=0.5\textwidth]{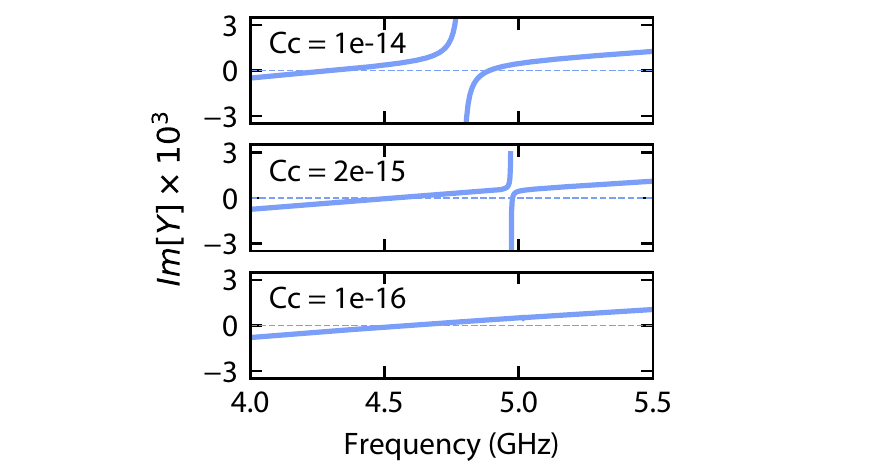}
\caption{
\textbf{Determining zero-point fluctuations from differentiating the admittance in decoupled circuits.}
Imaginary part of the admittance $\text{Im}[Y]$ across the junction of the circuit of Fig.~\ref{fig:1} for different values of the coupling capacitor $Cc$ and for $L_j=12$ nH.
Resonances correspond{} to the frequencies $\omega$ at which the admittance crosses 0, and the calculation of zero-point fluctuations depends on the derivative $\text{Im}[Y']$ at that point.
As the resonator and transmon parts of the circuit decouple, $\text{Im}[Y']$ becomes larger, requiring a lower $\delta\omega$ if the admittance is to be determined through Eq.~(\ref{eq:finite_difference_dY}).
In extreme cases (see lower panel), when the derivative is very large, the smaller variation in the background slope may be mistaken for the slope at a resonance.
}
\label{fig:dY_from_differentiation}
\end{figure}

But finding a good value of $\delta\omega$ is no easy task.
As an example, we consider the circuit of Fig.~\ref{fig:1}, where we have taken as reference element the junction.
As shown in Fig.~\ref{fig:dY_from_differentiation}, when the resonator and transmon decouple through a reduction of the coupling capacitor, a smaller and smaller $\delta\omega$ is required to obtain $Y_r'$ evaluated at the $\zeta_1$ of the resonator-like mode.
We have tried making use of Ridders method of polynomial extrapolation to try and reliably approach the limit $\delta x\rightarrow 0$~\cite{press2007numerical}.
However at small coupling capacitance, it always converges to the slower varying background slope of $Y_r$, without any way of detecting the error.

\subsubsection{Transfer functions from the admittance matrix}
Calculating the transfer function $T_{ij}$ could alternatively be carried out through the resolution of the system of equations (\ref{eq:admittance_matrix}).
The difference in voltage of a reference elements nodes would first have to be fixed to the zero-point fluctuations computed with the method of Sec.~(\ref{sec:methods_network_dY}).
These equations would have to be resolved at each change of system parameters and for each mode, with $\omega$ replaced in the admittance matrix by its corresponding value for a given mode.
This is to be balanced against a single symbolic derivation of $T_{ij}$ through star-mesh transformations, and fast evaluations of the symbolic expression for different parameters.

\section{Performance and limitations}
\subsection{Number of nodes}
In this section we ask the question: how big a circuit can QuCAT analyze?
To address this, we first consider the circuit of Fig.~\ref{fig:mmusc}(c), and secondly the same circuit with resistors added in parallel to each capacitor.
As the number of (R)LC oscillators representing the modes of a CPW resonator is increased, we measure the time necessary for the initialization of the Qcircuit object.
This is typically the most computationally expensive part of a QuCAT usage, limited by the speed of symbolic manipulations in Sympy.

These symbolic manipulations include:
\begin{itemize}
\item Calculating the determinant of the admittance matrix
\item Converting that determinant to a polynomial
\item Reducing networks through star-mesh transformations both for admittance and transfer function calculations
\item Rational function manipulations to prepare the admittance for differentiation
\end{itemize}
Once these operations have been performed, the most computationally expensive step in a Qcircuit method is finding the root of a polynomial (the determinant of the admittance matrix) which typically takes a few milliseconds.

\begin{figure}[ht!]
\includegraphics[width=0.5\textwidth]{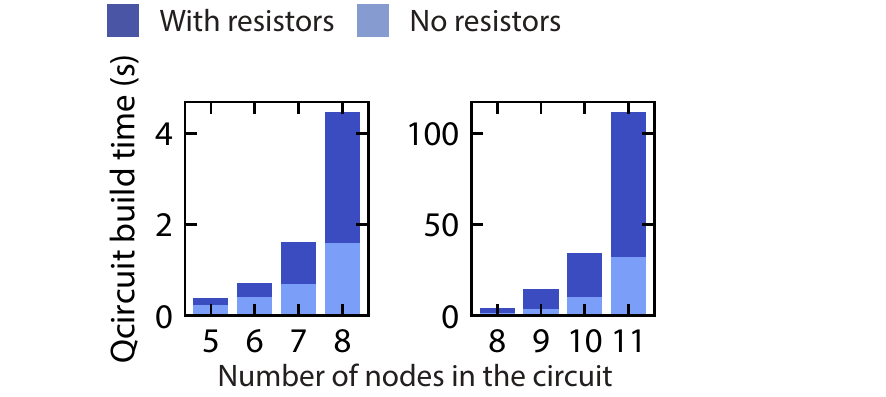}
\caption{
\textbf{Computation time with increasing circuit size. }
On the vertical axis, we show the time necessary to initialize the \inline{Qcircuit} object, which is the computationally expensive part of a typical QuCAT user case.
This is plotted as a function of the number of nodes in the circuit.
The test circuit used here is the multi-mode circuit of Fig.~\ref{fig:mmusc}(c), optionally with a resistor in parallel of each capacitor.
The number of nodes are increased by adding modes to the circuit.
Most of the computational time is spent in the symbolic manipulations performed with the sympy library.
}
\label{fig:performance}
\end{figure}

The results of this test are reported in Fig.~\ref{fig:performance}. 
We find that relatively long computation times above 10 seconds are required as one goes beyond 10 circuit nodes.
Due to an increased complexity of symbolic expressions, the computation time increases when resistors are included.
For example, the admittance matrix of a non-resistive circuit will have no coefficients proportional to $\omega$, only $\omega^2$ and only real parts, translating to a polynomial in $\Omega = \omega^2$ which will have half the number of terms as a resistive circuit.
However, we find that this initialization time is also greatly dependent on the circuit connectivity, and this test should be taken as only a rough guideline.
Making QuCAT compatible with the analysis of larger circuits will inevitably require the development of more efficient open-source symbolic manipulation tools.
The development of the open-source C++ library SymEngine \url{https://github.com/symengine/symengine}, together with its Python wrappers, the symengine.py project \url{https://github.com/symengine/symengine.py}, could lead to rapid progress in this direction.
An enticing prospect would then be able to analyze the large scale cQED systems underlying modern transmon-qubit-based quantum processors~\cite{2019APS..MARA42002K}.
One should keep in mind that an increase in circuit size translates to an increase in the number of degrees of freedom of the circuit and hence of the Hilbert space size needed for further analysis once a Hamiltonian has been extracted from QuCAT.

\subsection{Degree of anharmonicity}\label{sec:high_anharmonicity}
\begin{figure}[ht!]
\includegraphics[width=0.5\textwidth]{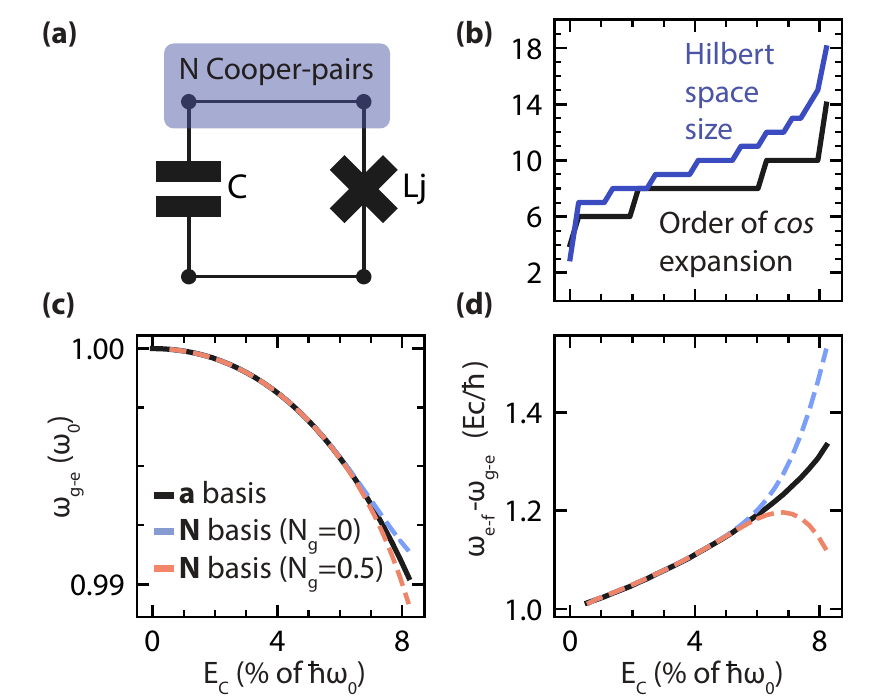}
\caption{
\textbf{Applicability of the harmonic Fock basis.}
\textbf{(a)} Transmon or Cooper-pair-box circuit.
%
%
\textbf{(b)} On the x-axis, we vary the approximate anharmonicity $E_C= e^2/2C$ with respect to the frequency $\omega_0=1/\sqrt{L_jC}$.
For each value, we plot a Hilbert space size, and order of Taylor-expansion of the junction cosine-potential.
Incrementing these values produces less than a 0.1 percent change in the first two transition frequencies obtained by diagonalizing the Hamiltonian.
Beyond a relative anharmonicity of 8, convergence is no longer reached, even for Hilbert space sizes and Taylor expansions up to 100.
\textbf{(c)}
Frequency of the first transition  $\omega_{g-e}$ obtained from Hamiltonian diagonalization, relative to the value expected from first order perturbation theory: $\omega_0-E_C/\hbar$.
\textbf{(d)}
Anharmonicity $\omega_{e-f}-\omega_{g-e}$ obtained from Hamiltonian diagonalization relative to the value expected from first order perturbation theory $E_C/\hbar$.
Black lines correspond to a diagonalization in the harmonic Fock basis (Eq.~(\ref{eq:hamiltonian_taylor_SI})), blue and orange dashed lines correspond to a diagonalization of the Cooper-pair box Hamiltonian with gate charges of $0$ and $1/2$ respectively.
The harmonic Fock basis provides reliable results up to approximatively 6 percent anharmonicity.
%
}
\label{fig:high_anharmonicity}
\end{figure}
In this section we study the limits of the current quantization method used in QuCAT.
More specifically, we study the applicability of the basis used to express the Hamiltonian, that of Fock-states of harmonic normal modes of the linearized circuit.
To do so, we use the simplest circuit possible (Fig.~\ref{fig:high_anharmonicity}(a)), the parallel connection of a Josephson junction and a capacitor.
As the anharmonicity of this circuit becomes a greater fraction of its linearized circuit resonance, the physics of the circuit goes from that of a Transmon to that of a Cooper-pair box~\cite{Koch2007}, and the Fock-state basis becomes inadequate.
This test should be viewed as a guideline for the maximum acceptable amount for  anharmonicity.
We find that when the anharmonicity exceeds 6 percent of the eigenfrequency, a QuCAT generated Hamiltonian will not reliably describe the system.

In this test, we vary the ratio of Josephson inductance $L_j$ to capacitance $C$, increasing the anharmonicity expected from first-order perturbation theory (see Eq.~\ref{eq:hamiltonian_first_order}), called charging energy $E_C = e^2/2C $.
The resonance frequency of the linearized circuit $\omega_0 = 1/\sqrt{L_jC}$ is maintained constant.
For each different charging energy, we use the \inline{hamiltonian} method to generate a Hamiltonian of the system. 
We are interested in the order of the Taylor expansion of the cosine potential, and the size of the Hilbert space, necessary to obtain realistic first and second transition frequencies of the circuit, named $\omega_{g-e}$ and $\omega_{e-f}$ respectively.
To do so, we increase the order of Taylor expansion, and for each order we sweep through increasing Hilbert space sizes.
In Fig.~\ref{fig:high_anharmonicity}(b), we show the values of these parameters at which incrementing them would not change $\omega_{g-e}$ and $\omega_{e-f}$ by more than 0.1 percent.
Beyond a relative anharmonicity $E_C/\hbar\omega_0$ of 8 percent, such convergence is no longer reached, even for cosine expansion orders and Hilbert space sizes up to 100.

Up to the point of no convergence, we compare the results obtained from the diagonalization in the harmonic Fock basis (Hamiltonian generated by QuCAT), with a diagonalization of the Cooper-pair box Hamiltonian.
In regimes of higher anharmonicity, the system becomes sensitive to the preferred charge offset between the two plates of the capacitor $N_g$ (expressed in units of Cooper-pair charge $2e$) imposed by the electric environment of the system.
The Cooper-pair box Hamiltonian takes this into account
\begin{equation}
\begin{split}
    \hat H_\text{CPB} = 4E_C(\sum_N\ket{N}\bra{N}-N_g)^2\\-\sum_NE_j(\ket{N+1}\bra{N}+\ket{N}\bra{N+1})
\end{split}
\end{equation}
where $\ket{N}$ is the quantum state of the system where N Cooper-pairs have tunneled across the junction to the node indicated in Fig.~\ref{fig:high_anharmonicity}(a).
For more details on Cooper-pair box physics and the derivation of this Hamiltonian, refer to Ref.~\cite{schuster2007circuit}.
We diagonalize this Hamiltonian in a basis of 41 $\ket{N}$ states.
We find that beyond 6 percent anharmonicity, the Cooper-pair box Hamiltonian becomes appreciably sensitive to $N_g$ and diverges from the results obtained in the Fock basis.
This corresponds to $E_j/E_C \simeq 35$ at which the charge dispersion (the difference in frequency between $0$ and $0.5$ charge offset) is $4\times10^{-5}$ and $1\times10^{-3}$ for the first two transitions respectively.

Beyond 8 percent anharmonicity, one cannot reach convergence with the Fock basis and just before results diverge considerably from that of the Cooper-pair box Hamiltonian.
This corresponds to $E_j/E_C \simeq 20$ at which the charge dispersion is $1.5\times10^{-3}$ and $3\times10^{-2}$ for the first two transitions respectively.
A possible extension of the QuCAT Hamiltonian could thus include handling static offsets in charge and different quantization methods, for example quantization in the charge basis to extend QuCAT beyond the scope of weakly-anharmonic circuits.

\section{Installing QuCAT and dependencies}
The recommended way of installing QuCAT is through the standard Python package installer by running \inline{pip install qucat} in a terminal.
Alternatively, all versions of QuCAT, including the version currently under-development is available on github at \url{https://github.com/qucat}.
After downloading or cloning the repository, one can navigate to the \inline{src} folder and run \inline{pip install .} in a terminal.

QuCAT and its GUI is cross-platform, and should function on Linux, MAC OS and Windows.
QuCAT requires a version of Python 3, using the latest version is advised.
QuCAT relies on several open-source Python libraries: Numpy, Scipy, Matplotlib, Sympy and QuTiP~\cite{johansson2012qutip,johansson2013qutip}, installation of Python and these libraries through Anaconda is recommended.
The performance of Sympy calculations can be improved by installing Gmpy2.
\section{List of QuCAT objects and methods}
\setlength\parindent{0pt}    

\textbf{QuCAT objects }

\inline{Network} -- Creates a \inline{Qcircuit} from a list of components

\inline{GUI}  -- Opens a graphical user interface for the construction of a \inline{Qcircuit}

\inline{J}  -- Creates a Josephson junction object

\inline{L} -- Creates a inductor object

\inline{C} -- Creates a capacitor object

\inline{R} -- Creates a resistor object

\vspace{5pt}
\textbf{Qcircuit methods}

\inline{eigenfrequencies}  -- Returns the normal mode frequencies 

\inline{loss_rates}  -- Returns the normal mode loss rates

\inline{anharmonicities}  -- Returns the anharmonicities or self-Kerr of each normal mode 

\inline{kerr} -- Returns the self-Kerr and cross-Kerr for and between each normal mode  

\inline{f_k_A_chi} -- Returns the eigenfrequency, loss-rates,  anharmonicity, and Kerr parameters of the circuit  

\inline{hamiltonian} -- Returns the Hamiltonian of Ref.~\ref{eq:hamiltonian_taylor_SI}  

\vspace{5pt}
\textbf{Qcircuit methods}  \textit{(only if built with GUI)}  

\inline{show}  -- Plots the circuit 

\inline{show_normal_mode}  -- Plots the circuit overlaid with the currents,  voltages, charge or fluxes through each component when a normal mode is populated with a single-photon coherent state  

\vspace{5pt}
\textbf{J,L,R,C methods}

\inline{zpf}  -- Returns contribution of a mode to the zero-point fluctuations  in current, voltages, charge or fluxes 

\vspace{5pt}
\textbf{J methods}

\inline{anharmonicity}  -- Returns the contribution of this junction to the anharmonicity of a given normal mode (Eq.~(\ref{eq:Amj}))

\vspace{5pt}

\section{Source code and documentation}
The code used to generate the figures of this paper are available in Zenodo with the identifier 10.5281/zenodo.3298107.
Tutorials and examples, including those presented here are available on the QuCAT website at \url{https://qucat.org/}.
The latest version of the QuCAT source code, is available to download or to contribute to at \url{https://github.com/qucat}.

\section{Acknowledgments}
We acknowledge Marios Kounalakis for useful discussions and for reading the manuscript.
This work was supported by the European Research Council under the European Union’s H2020 program [grant numbers 681476-QOM3D, 732894-HOT, 828826-Quromorphic].
\section{Author contributions}
MFG developed QuCAT and the underlying theory and methods. 
MFG wrote this manuscript with input from GAS. 
GAS supervised the project.

\vspace{5pt}
\textit{Declarations of interest: none.}
\bibliography{library}

\end{document}